\documentclass[fleqn,usenatbib]{mnras}

\usepackage{newtxtext,newtxmath}

\usepackage[T1]{fontenc}

\DeclareRobustCommand{\VAN}[3]{#2}
\let\VANthebibliography\thebibliography
\def\thebibliography{\DeclareRobustCommand{\VAN}[3]{##3}\VANthebibliography}

\usepackage{graphicx}	
\usepackage{amsmath}	

\usepackage{amssymb}	

\usepackage{hyperref}
\hypersetup{
    colorlinks=true,
    linkcolor=blue,
    filecolor=magenta,      
    urlcolor=cyan,
}

\title[Detecting sub-pc SMBBHs: long-term monitoring perspective]{Detecting subparsec super-massive binary black holes: Long term monitoring perspective}

\author[L. \v C. Popovi\'c et al.]{
L. \v C. Popovi\'c,$^{1,2,3}$\thanks{E-mail: lpopovic@aob.rs}
S. Simi\'c,$^{4}$
A. Kova\v cevi\'c,$^{2,3}$
D. Ili\'c$^{2,5}$
\\
$^{1}$Astronomical Observatory, Volgina 7, 11000 Belgrade, Serbia\\
$^{2}$Department of astronomy, Faculty of mathematics, University of Belgrade, Studentski trg 16, 11000 Belgrade, Serbia\\
$^{3}$ Fellow of Chinese Academy of Sciences President's International
Fellowship Initiative (PIFI) for visiting scientist\\
$^{4}$Faculty of Science, University of Kragujevac, Radoja Domanovi\'ca 12, 34000 Kragujevac, Serbia\\
$^{5}$ Humboldt Research Fellow, Hamburger Sternwarte, Universit{\"a}t Hamburg, Gojenbergsweg 112, 21029 Hamburg, Germany\\
}

\date{Accepted XXX. Received YYY; in original form ZZZ}

\pubyear{2020}

\begin{document}
\label{firstpage}
\pagerange{\pageref{firstpage}--\pageref{lastpage}}
\maketitle

\begin{abstract}
Here we consider the perspective to detect sub-pc super-massive binary
black-hole (SMBBH) systems using long-term photometric and spectroscopic
monitoring campaigns of active galactic nuclei. This work explores the
nature of long-term spectral variability caused by the dynamical effects
of SMBBH systems. We describe in great detail a model of SMBBH system
which considers that both black holes have their accretion disc and
additional line emitting region(s).
We simulate the H$\beta$ spectral band (continuum+broad H$\beta$ line)
for different mass ratios of components and different
total masses of the SMBBH systems ($10^6-10^8\mathrm{M\odot}$). We analyze the
set of continuum and broad line light curves for several full orbits of
SMBBHs with different parameters, to test the possibility to extract the
periodicity of the
system. We consider different levels of the signal-to-noise ratio, which
is added to the simulated spectra. Our analysis showed
that the continuum and broad line profiles emitted from an SMBBH system
are strongly dependent, not only on the mass ratio of
the components but also on the total mass of the system. We found that
the mean broad line profile and its rms could indicate the
presence of an SMBBH. However, some effects caused by the dynamics
of a binary system could be hidden due to a low signal-to-noise ratio.
Finally, we can conclude that the long-term AGN monitoring campaigns
could be beneficial for the detection of SMBBH candidates.
\end{abstract}




\section{Introduction}

It is well known that there are many galaxy mergers where a collision occurs on kpc scale \citep[see e.g.][etc.]{bu11,fu11,ro13,Woo13,Woo14,Liu14,shan16,liu18}.
Eventually, a two colliding galaxies can evolve into so-called sub-pc phase, where two super-massive black holes (SMBHs), surrounded by local gas and stellar host, build a super-massive binary black hole (SMBBH) system. This phase is a prior SMBBH stage of the coalescence that is expected to be a source of gravitational waves (GWs). Thus, as sub-pc SMBBHs are a potential source of GWs, their detection and location on the sky are essential, especially after recent detection of the gravitational-wave signals \citep[see e.g., for GW150914 in][]{abb16}.
Additionally, future superior radio facilities such as Five-hundred-meter Aperture Spherical Telescope (FAST) and Square Kilometre Array (SKA) are going to improve the sensitivity of Pulsar Timing Array (PTA) project for
GW detection \citep[][]{lam18,nguyen20}. The GW sources for PTA detection are SMBBHs, therefore finding a sample of SMBBH candidates may be very significant for future PTA GW detection.

The SMBBH system is expected to reside in the center of a number of
galaxies \citep{Begelman80,Merritt05}, however, there is a problem in detecting SMBBHs. Direct imaging in the high resolution radio observations
could be potentially a good method for SMBBH detection \citep[see e.g.][etc.]{bu11,fu11,ro13,ts13,Liu14,mo18}, especially on kpc scale distances  \citep[see][]{fu11}.
At sub-pc scale, an SMBBH system usually cannot be spatially resolved by present day available observational facilities, especially for the high-redshift objects\footnote{For the objects in the local universe, exceptions are the Event Horizon Telescope (https://eventhorizontelescope.org/) and Gravity (https://www.eso.org/sci/facilities/paranal/instruments/gravity.html), which can observe with a very high spatial resolution.}. But, in the case where the
spectral characteristics reflect the dynamical signature of an orbital motion of the binary system, spectral observations can be used for detection of the SMBBH candidates \citep[see e.g.][etc.]{ga83,pop00,sl10,ts11,er12,pop12,bon12,Liu14,ba15,nb16,li16,sp16,wa17,Li19,derosa19,nguyen20,ser20}.

Typically, a sub-pc SMBBH system that is surrounded by gas produces an activity that is similar to the one observed in an active galactic nucleus (AGN).
This has an advantage that we could detect the activity produced by an
SMBBH system, but also it has a disadvantage that it is hard to separate
an SMBBH activity from the single AGN activity  \citep[see for a review][]{pop12}.
For example, some peculiar spectral line profiles can be explained with both
single and SMBBH models (as e.g. double peaked emission lines).  Also, the
periodic oscillation which should be present in the light curves are often
hidden by the stochastic  nature of the AGN activity and different
observational effects (e.g. poor  S/N ratio, cadence, etc.).

There are several spectroscopic features that indicate a presence of sub-pc SMBBHs
\citep[see e.g.][etc.]{er12,pop12,bon12,gr15,gr15a,li16,gu19,ser20}
but, in addition, there is a great deal of effects \citep[as e.g. random fluctuations caused by explosion or giant flares, see][]{Ci18,gr18}. These can affect the observed spectra (continuum and spectral lines) and hide the signatures of an SMBBH presence which are expected to be seen, e.g. in periodicity, and/or line profiles variations.

To detect sub-pc SMBBHs, the large spectroscopic surveys have been used, as e.g. SDSS \citep[Sloan Digital Sky Survey, see e.g.][]{ts11,er12,ju13,Liu14,gr15}, that can give some indications for a number of objects which are good SMBBH candidates. To confirm an SMBBH presence one should follow the spectral variability of the system \citep{bon12,ru15,ba15,sh16,li16,ko17,kov19,kov20}.
It is expected that SMBBHs have similar variability  \citep{ru15,sh16,ko17,kov19,kov20},
but with some particular characteristics caused by dynamical effects of a binary system \citep[as e.g. periodicity,see][]{bon12,gr15,gr15a,li16,bon16,ko17,kov19,kov20,Li19,derosa19} that can be used to distinguish it from others mechanisms which produce spectral variability.

Moreover, in the next decade, we expect to have comprehensive AGN monitoring
campaigns, out of which the Vera C. Rubin Observatory Legacy Survey of Space and Time (LSST) with its very high cadence 10-year survey seems to be very promising for the detection of SMBBH candidates \cite[see, e.g.][]{br18}. Futhermore, the Maunakea Spectroscopic Explorer (MSE) with its multi-object spectrographs will provide the spectral survey of a large number of AGNs, and complement the photometric surveys in the quest for SMBBH systems \citep[see][]{mse19}. These motivated us to investigate in time-domain the influence of the dynamical effect of a binary system to
a spectral energy distribution (SED) and broad line profiles of an SMBBH system. By knowing those details, one could improve the detection probabilities of SMBBH systems and shape future monitoring campaigns.

In this paper, we elaborate on the phenomenological model of an SMBBH
system, introducing  complex structure of the continuum and 
line-emitting regions, and exploring the time-domain perspective
necessary in the analysis of the long-term monitoring results. We calculate the resulting SED of the SMBBH system, and explore the variability in the continuum at 5100 \AA\ and H$\beta$ broad line luminosity, as well as in the broad line profile. These spectral features are selected as they are usually observed in the spectral monitoring campaigns, but are also covered in photometric surveys. Beside dynamical effects, we analyze the influence of the total mass of the system and the mass ratio of the components to the resulting spectra, as well as the effect of adding the white noise to the simulated SED. We aim to determine properties of correlations between continuum and emission line light curves of an SMBBH system, and to test the possibility to detect periodicity in these light curves, which is important for the future long-term time-domain surveys.

The paper is organized as follows. In \S \ref{sec:model} we describe in great details the model and used assumptions, and in \S \ref{sec:results} we present results of the analysis of the simulated SED (continuum and broad H$\beta$ emission line), with the emphasis on the variability of continuum and line emission.
Finally in \S \ref{sec:conclusion} we outline our conclusions.

\section{The SMBBH model}
\label{sec:model}

An AGN hosts an SMBH in the center with an accretion disc, which continuum
emission ionizes the gas in the broad line region (BLR). The BLR consists of a large number of emitting clouds and, in a whole, is optical thin \citep[see][]{ga09}, i.e., there are no effects of radiative transfer. The BLR seems to be flattened with the inclination that is similar to the accretion disc (and dusty torus) inclination \citep[][]{co06, sa19}.

The model considers two AGNs at a sub-pc mutual distance in which both SMBHs have their own accretion disc and BLR, and consequently, that activity can be detected from both SMBHs, i.e., both components are emitting the electromagnetic spectrum that is typical for an AGN, similarly as in \cite{sp16}. In addition, in this paper we introduce that SMBBH system is surrounded by a common circum-binary BLR (for more details in \S 2.4).

\begin{figure*}
\centering
\hskip -4mm
\includegraphics[width=16.5cm]{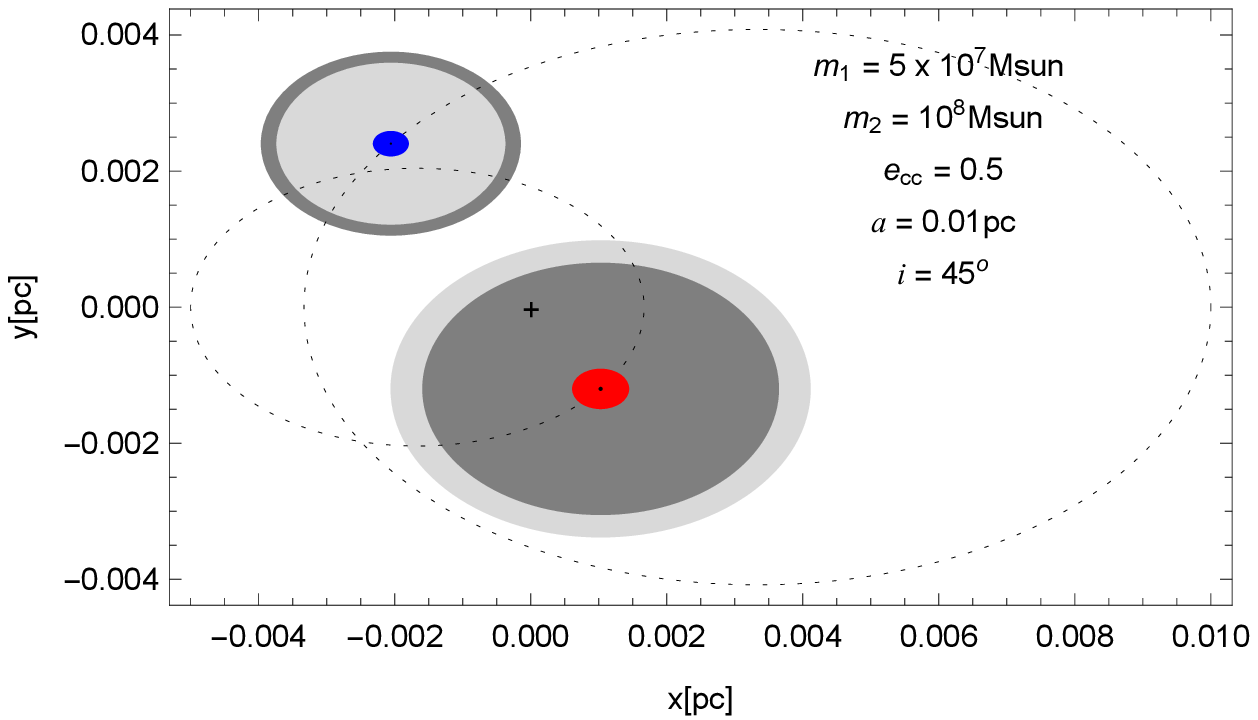}\\
\hskip 14mm
\includegraphics[width=14cm]{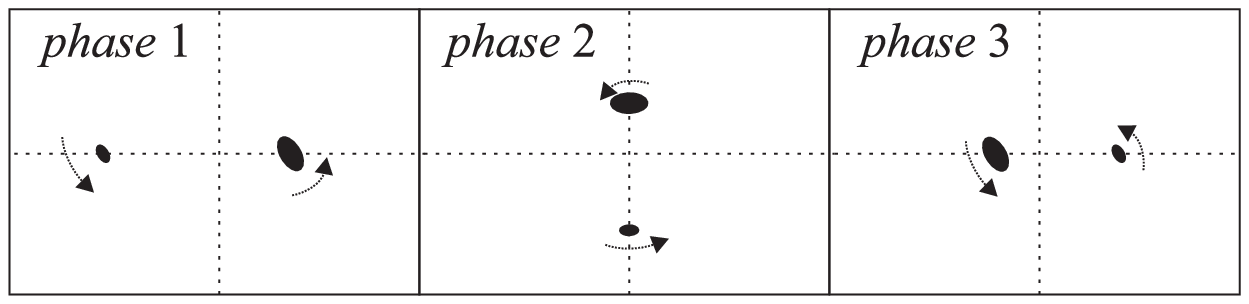}
\caption{One example of the SMBBH model. Accretion disc surrounding each SMBH (small black dot in the center) is denoted with blue (lower SMBH mass $m_1$) and red disc (higher SMBH mass $m_2$). Dark grey region depicts the BLR outer limit defined with the Roche lobe, and light gray region defines the BLR outer limit empirically calculated (see \S \ref{sec:kinematicsBLR} for details). Dashed ellipses are orbits of SMBHs, and the x and y axis give the position in pc with the respect to the SMMBH barycenter, denoted with cross. Model input parameters (SMBH masses $m_1$, $m_2$, eccentricity $e_{cc}$, mean distance between components $a$, orbital inclination $i$) are given in the plot. Bottom panel shows three different phases defined with $t=0$, $t=P_{\mathrm{orb}}/4$ and $t=P_{\mathrm{orb}}/2$, respectively from left to right. The assumed orbital inclination of $i=45^o$ is kept constant.}
\label{fig:SMBBH_sketch}
\end{figure*}

Besides standard Newtonian consideration of dynamical effects of SMBBH
system on spectral observables we introduced in our model additional features discussed in further subsections. Fig. \ref{fig:SMBBH_sketch} (upper panel) illustrates the assumed SMBBH system in which each SMBH has its own accretion disc, which is a continuum source, and the BLR, which emits broad lines.

To investigate the variability of the continuum and line emission, and study the periodicity in the light
curves, we introduce  a simple dynamical model in which the SMBBH gravitationally interacts with the accretion disc surrounding the opposite SMBH, affecting the surface disc temperature. Consequently, this disc accretion rate is altered, causing the change in the continuum and line emission.

We discuss separately different aspects of the proposed model: 1) the dynamical parameters of the SMBBH system; 2) the structure of the disc that is the source of the continuum emission; 3) the sources of the continuum variability; 4) the structure of the BLRs in both components; and 5) the empirical constraints in the broad line parameters defined by dynamical and physical properties of the central source.

\subsection{Dynamical parameters of the SMBBH system}
\label{sec:dyn_par}

Our model of the binary system contains two SMBHs, which one realization is shown in Fig. \ref{fig:SMBBH_sketch}. We discuss the most general case where components have arbitrary masses designated in our computation as $m_1$ and $m_2$, and the ratio of component masses is defined as \begin{equation*}q=\frac{m_1}{m_2},\end{equation*} with 
$m_1 < m_2$.
Orbits in such a system are elliptical with eccentricity $e_{cc}$.

In general equation for the period P in binary system can be computed from the third Kepler law, as \citep{Hil01}:
\begin{equation}
P^2=\frac{4\pi^2a^3}{G(m_1+m_2)},
\label{eq:period}
\end{equation}
where $a$ is a mean distance between components, $G$ universal gravitational constant.

In our particular case of SMBBHs, the orbital period $P_{\mathrm{orb}}$ can be written as:
\begin{equation}
P_{\mathrm{orb}}=210\left(\frac{a}{0.1pc}\right)^{3/2}\left(\frac{2\times10^8 \mathrm{M\odot}}{m_1+m_2}\right)^{1/2}
\label{eq:period1}
\end{equation}
that is given in years. The mean distance between components $a$ is given in pc and masses $m_{1,2}$ in $10^8\mathrm{M\odot}$.

To compute orbital motion and radial velocities of components, we use the standard Keplerian approach. We introduce mean anomaly as
\begin{equation*}
M=\frac{2\pi}{P_{\mathrm{orb}}}(t-\tau)=2\pi\Phi
\end{equation*} where $\tau$ is initial moment of measurements, $t$ is time variable and $\Phi$ is orbital phase of the system,
usually ranging from 0 to 1.
Using the Kepler's equation given in a form of:
\begin{equation}
M=E-e_{cc}\sin(E)
\label{eq:kepler}
\end{equation}
we can numerically extract eccentric anomaly $E$, which than can be used to compute true anomaly as:

\begin{equation}
\theta=2\arctan\left(\sqrt{\frac{1+e_{cc}}{1-e_{cc}}}\tan\frac{E}{2}\right)
\label{eq:true_anomaly}
\end{equation}

Now, the orbital paths or true barycentric orbits of the SMBBHs in the system are given by:
\begin{equation}
r_{1,2}(\theta)=\frac{a_{1,2}(1-e_{cc}^2)}{1+e_{cc}\cos(\theta)}
\label{eq:paths}
\end{equation}
with $a_{1,2}$ computed from the condition of placing origin of coordinate system in the  barycentre, which gives:

\begin{equation}
a_1=\frac{q a}{1+q} \quad \mathrm{and} \quad a_2=\frac{a}{1+q}
\end{equation}
In computations we assume that components differ in orientation by $180^\mathrm{o}$ or   $0.5\mathrm{P_{orb}}$. SMBHs barycentric positions are used for the calculation of continuum luminosity variability due to the disc temperature perturbation (see \S 2.3 and Appendix A).

The inclination of the orbital plane $i$ is  an angle between the normal on the orbital plane and the line-of-sight.
In the case when this angle is zero, the radial velocity in the observer frame will also be zero. In principle, the radial velocity
cannot be observed in the face-on orientation of an SMBBH system. Thus in our model, we assumed an orbital inclination angle to be $45^{\circ}$. As the orbit is elliptic, also  the azimuthal angle is important, but for a system it will be a constant for an observer. In this paper we take it to be equal to zero. In further research we plan to investigate the influence of azimuthal angle in our model in more detail.
This configuration is presented in Fig. \ref{fig:SMBBH_sketch}.

Radial velocities of components and their semi-amplitudes can be computed in the observer frame using:
\begin{equation}
v_{1,2}^\mathrm{rad}(\theta)=K_{1,2}\left[\cos(\theta +\omega)+e_{cc} \cos(\omega)\right]+\gamma, 
\label{eq:rad_velocities}
\end{equation}

\begin{equation}
K_{1,2}=\frac{2\pi a_{1,2}\sin(i)}{P_\mathrm{orb}\sqrt{1-e_{cc}^2}}
\label{eq:rad_velocities1}
\end{equation}
where $\omega$ is argument of perihelion which can have arbitrary values, but we choose $\omega=30^o$ in order to examine most interesting case and $\gamma$ is systemic velocity which in our case is equal to 0.

Effectively, SMBHs radial velocities are used for the calculation of continuum emission (\S 2.2), luminosity variability (\S 2.3) and for defining physical properties of BLRs (\S 2.4).
The bottom panel of Fig. \ref{fig:SMBBH_sketch} shows the scheme of three dynamical phases ($t=0$, $t=P_{\mathrm{orb}}/4$ and $t=P_{\mathrm{orb}}/2$ respectively from left to right), i.e., the position during the system revolution defined by time and angular velocity of SMBBH.

\subsection{The structure of the accretion disc -- continuum emission}

In our model both components have the accretion disc and the BLR surrounding the discs, as shown in Fig. \ref{fig:SMBBH_sketch}. For calculating the disc continuum emission in the UV-optical-IR bands, we use the model of a standard optically thick, geometrically thin, black body accretion disc \citep[see e.g.][]{Pringle72, Shakura73, Novikov73}, which effective temperature $T^i_{eff}$ (for $i=1,2$) as a function of radius from the center is given as \citep{Yan14}:
\begin{equation}
T^i_{\mathrm{eff}}[K]=2\cdot10^5 \left(\frac{10^8}{m_i}\right)^{1/4} \left(\frac{R_{\mathrm{in}}}{R}\right)^{\beta}\left(1-\sqrt{\frac{R_{\mathrm{in}}}{R}}\right)^{1/4}\left(\frac{0.1}{\epsilon}\right)^{1/4}
\left(\frac{f_E}{0.3}\right)^{1/4}
\label{eq:TodR}
\end{equation}
where $R_{\mathrm{in}}$ is the inner disc radius, $m_i$ is the SMBH mass of the $i$-th component of an SMBBH, and $\beta$ power law
index equal to 3/4 in the standard disc model, although different values can be considered.
$\epsilon$ is the radiative efficiency, while  $f_E$ is the Eddington ratio that is: 
$$
f_E = {\dot{M}_{\rm acc}\over\dot{M}_{\rm Edd}}, 
$$
where $\dot{M}_{\rm acc} $ is the accretion rate and $\dot{M}_{\rm Edd} $ is the Eddington accretion rate. In the  single SMBH case $f_E$ is usually assumed to be $\sim0.3$.

For a disc part that is not close to the central SMBH, Eq. \ref{eq:TodR} simplifies to $T\propto R^{-\beta}$.
We consider the emission from the UV to IR spectral range. Therefore we use only thermal emission as the primary radiation mechanism, and neglect the contribution of the inverse Compton and synchrotron radiation, as they are significant in the X-ray and $\gamma$ bands.

In such case, the radiated power emitted by a small ring-like
element of the disc surface $dS=2\pi rdr$ at the distance $r$ from the system center, with effective temperature $T_{\mathrm{eff}}$ defined in Eq. \ref{eq:TodR} is given as \citep{Poindexter08}:
\begin{equation}
dL= 4\pi\frac{2 h c^2}{\lambda^5}\frac{dS \cos(i)}{\left[\exp(\frac{hc}{\lambda k_{B}T_{\mathrm{eff}}})-1\right]}
\label{eq:dL}
\end{equation}
where $h,c,k_\mathrm{B}$ are Plank constant, speed of light, and Boltzman constant, respectively. $\theta_{\mathrm{disc}}$ is the inclination of the disc of one SMBH component.
To compute total luminosity of a disc (i.e., the SED), we integrate Eq.
\ref{eq:dL} over entire surface of the disc and obtain:
\begin{equation}
L(\lambda)\propto \int_{S_{\mathrm{disc}}}\lambda dL(\lambda,T_{\mathrm{eff}})
\label{eq:SED}
\end{equation}

For the above equation we need to define the inner $R^{\mathrm{in}}_i$ and outer $R^{\mathrm{out}}_i$ radii of the disc.
For the inner disc radius we adopt $R^{\mathrm{in}}_i\sim 10\ R_\mathrm{g}$, since we focus only on the UV/optical/IR
emission. For the outer disc radius, we use the relationship given by  \cite{Vicente14}, in which the outer radius, in units of light days (ld), is defined as:
\begin{equation}
R^{\mathrm{out}}_i = \frac{1}{2}\cdot r_0\left[\frac{m_i[{\mathrm{M\odot}}]}{10^9}\right]^{2/3}
\label{eq:rho_out}
\end{equation}
where $i$=1,2 denotes the component of an SMBBH, and $r_0$ is $4.5^{+0.7}_{-1.6}$ ld. The exponent $2/3$ in Eq. \ref{eq:rho_out}, is expected in the case of Shakura-Sunayev accretion disc \citep[][]{Shakura73},
that is in agreement with estimates obtained by microlensing \citep[see e.g.][]{Morgan10}.

\begin{figure}
\includegraphics[width=8.8cm]{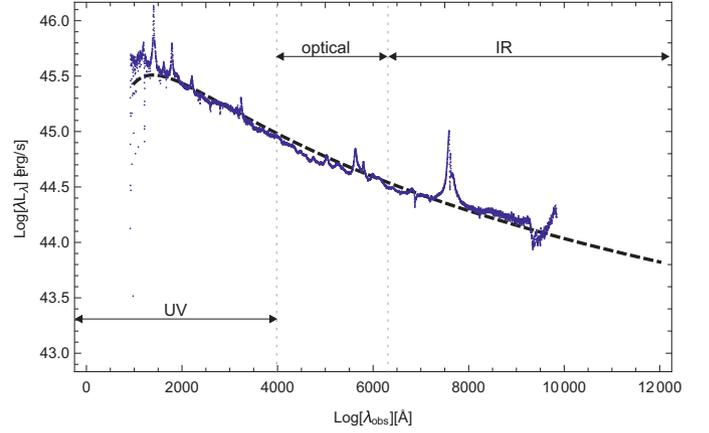}
\caption{An example of a real quasar spectrum described with our model. The observed spectrum of 3C273 taken from  \citet{Shang05} (blue dots) is simulated with our model of a single SMBH with an accretion disc (dashed black line).}
\label{fig:SEDfit3C273}
\end{figure}

To test our model for a case of a single SMBH with an accretion disc, we compare it with the observed spectrum of 3C273, as this is one of the most studied objects with high quality observations (Fig. \ref{fig:SEDfit3C273}). The modeled SED follows very well the observed spectrum of 3C273 in the UV, optical, and IR wavelength bands taken from \cite{Shang05}. The assumed parameters for the simulated model are: the accretion disc inclination $\theta_{\mathrm{disc}}=45^{\circ}$, and the SMBH mass of $m\approx2\times10^9\mathrm{M}_{\odot}$. The SMBH mass is a bit smaller then  obtained in \citep{pal05}, which is probably caused by the fixed inclination in our model.

We assume the optical continuum at $\lambda$5100 \AA\, reflects the ionizing continuum, which is important for the BLR dimension estimates and could be perturbed by interactions of SMBHs. In our analysis, we consider only the disc contribution to the ionizing continuum emission, but we note that other sources of the optical continuum could be present, as e.g., synchrotron optical continuum radiation from a jet.
However, the jet continuum radiation probably cannot ionize the BLR and is not relevant for this model.

\subsection{Luminosity variability due to dynamical interaction}
\label{sec:interaction_var}

First, in Eq. \ref{eq:dL} we include dynamical effects of radial motion taking the shift in the emitted energy
(wavelengths) of photons as:
$$ \lambda = \lambda_0(1+z_{\mathrm{dopp}}^i),$$
where $z^i_{\mathrm{dopp}}=v_i/c$. The velocity $v_i$ is calculated using Eq. \ref{eq:rad_velocities} and \ref{eq:rad_velocities1}. This dynamical effect of the radial motion of the SMBBH components is included in the line shift, but also in the continuum.

Second,  one should consider the gravitational effect, i.e., the perturbation of the accretion disc of one SMBBH component due to the interaction
caused by the companion. This interaction
depends mostly on the distance between components and
their masses. The analysis of the theory of disc perturbation due to tidal interaction is out of the scope of this work. We only consider that the gravitational interaction can affect the accretion rate of the disc, and consequently, the disc temperature profile of the components.

Due to the gravitational interaction of two SMBBH components (designated with indexes $i$ and $j$), the effective disc surface temperature of the $i$-th component can be modified as (see also Appendix A):
\begin{equation}
T_{\mathrm{eff}}^{i}=T_{o}^{i}\left(1+\frac{m_j}{m_i}\frac{R_i}{r_j(t)}\cdot\cos\beta\right)^{1/4},
\label{eq:Teff_per}
\end{equation}
where $T_{o}^{i}$ is the effective disc surface temperature for non-perturbed disc, $r_j(t)$ is the distance between
the perturbing SMBH (i.e., $j$-th component) and the part of the perturbed disc located at radius $R_i$ from the host SMBH (see illustration in Appendix A).
The distance $r_j(t)$ can be written as:
\begin{equation}
r_j(t)=a(t)\sqrt{1+\left(\frac{R_i}{a(t)}\right)^2-2\frac{R_i}{a(t)}\cdot\cos(\varphi)},
\label{eq:r_od_fi}
\end{equation}
\noindent where $a(t)$ is distance between two SMBHs, and $\varphi$ is the angle between  $a(t)$ and $R_i$ observed from the center of component $i$. For more details see explanation in Appendix A.

Third, the accretion rate of both SMBH components is affected by the binary dynamics that forms spiral streams which  accumulate additional matter near each SMBH \citep[forming so called “mini discs” of gas orbiting the individual SMBHs,][]{fa14}. This changes the accretion rates in both components. To include the changes in the accretion rates we applied the relations given in \cite{fa14} and find that the changes in effective temperature of $i$-th component has a form (see Appendix B):

\begin{equation}
T^i_\mathrm{eff}(t)=T^i_{o}\left(1+\frac{m_j}{m_i}\frac{R\cdot\cos\beta}{r(t)} \right)^{1/4} \cdot \left(\frac{f^i_E(t)}{f^{i0}_E}\right)^{1/4},
\label{eq:Teff_pert21}
\end{equation}
where the tidal perturbation and accretion rate changes are taken into account.

\begin{figure}
\centering
\includegraphics[width=8.7cm]{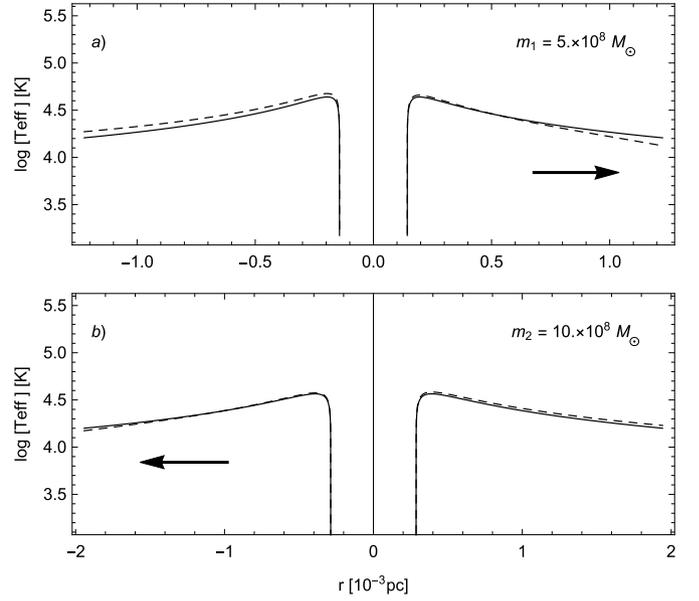}
\caption{Logarithm of the temperature profiles across the disc along the axis between the two interacting SMBHs at the selected distance of 0.01 pc with the masses of $0.5\times10^9\mathrm{M\odot}$ and $10^9\mathrm{M\odot}$ and BH minimum separation. The upper panel shows the temperature profile of the
disc around the SMBH of $0.5\times10^9\mathrm{M\odot}$ and the bottom of $10^9\mathrm{M\odot}$. Solid line represents the unperturbed and dashed the perturbed temperature profile. The arrows show the direction of the interaction, i.e., the position of the companion.}
\label{fig:Teff_profile_perihel}
\end{figure}

\begin{figure}
\centering
\includegraphics[width=8.7cm]{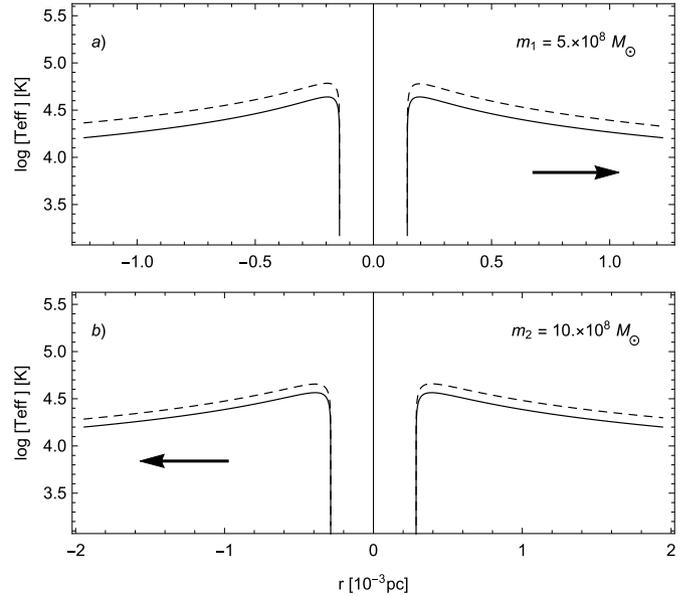}
\caption{Same as in Fig.\ref{fig:Teff_profile_perihel}, but for maximal BHs separation.}
\label{fig:Teff_profile_aphel}
\end{figure}

Eqs. \ref{eq:Teff_per},  \ref{eq:r_od_fi} and \ref{eq:Teff_pert21} show that the perturbation in the effective temperature is a function of the distance between components, which is time dependent. The amplitude of the temperature perturbation
also depends on the masses of components, and as shown in Fig. \ref{fig:Teff_profile_perihel}, the temperature perturbation will be stronger in the part of the disc closer to the perturbing component. As expected, closer to the perturbing SMBH (with lower mass $m_1$), the effective temperature in the disc is decreased by the perturbation. This is caused by the additional gravitational effect, which perturbs Keplerian-like rotation in this part.
On the other side, in the case of maximal SMBH separation, the accretion is more effective (see Fig. \ref{fig:Teff_profile_aphel}), both discs become brighter, and the total luminosity of the system is in the maximum.

Taking into account the interactions, and that  luminosity of the system is determined by the temperatures of both SMBH disc, we calculated the total luminosity, $L_\mathrm{tot}$, of the an SMBBH as:

\begin{equation}
L_\textrm{tot}(\lambda,t)\propto \int_{S_{\mathrm{disc}}}\lambda dL_\textrm{tot}(\lambda,T_{\mathrm{eff}},t)
\label{eq:SED1}
\end{equation}
where $dL_\mathrm{tot}$ is calculated, taking both ($dL_1$ and $dL_2$) components, as:
\begin{equation}
dL_\mathrm{tot}=dL_1+dL_2= 4\pi\frac{2 h c^2}{\lambda^5}
\left[\sum_{n=1,2}\frac{dS_n \cos(i_n)}{\left[\exp(\frac{hc}{\lambda k_{B}T^n_{\mathrm{eff}}(t)})-1\right]}\right].
\label{eq:dL1}
\end{equation}
$T^n_{\mathrm{eff}}(t)$ depends on orbital elements (as it is shown in Eq.  \ref{eq:Teff_pert21}), i.e., the total luminosity of the system is time dependent.

We note that the eclipsing effect, in which one of the components is eclipsing another, may cause the variability in the total luminosity of an SMBBH system. However, the eclipse could happen only in a near edge-on binary orientation; thus it is implausible. In addition, the accretion discs considered here are very thin i.e., the thickness of the disc is much smaller compared to the overall radius, so even if the eclipse occurs, we do not expect significant changes in the total emission of an SMBBH system. In our simulations, we fix the inclination of the binary rotation plane to be $i=45^{\circ}$, consequently, we will not consider this effect.

The additional dynamical effect we consider here is the variability of the observed emission caused by the Doppler beaming \citep{do15}. The luminosity of an SMBBH system is converted to the flux that an observer can detect. The apparent flux $F_\lambda$ at a fixed observed wavelength
$\lambda$ is modified from the flux of a stationary source $F_\lambda^{\rm 0}$. Following \cite{do15} one can obtain:
\begin{equation}
{\Delta F_\nu \over F_\nu} = {\Delta F_\lambda \over F_\lambda}= (3-\alpha)(v_i(t)/c),
\label{eq:dF_F}
\end{equation}
where $v_i(t)$ is the velocity of a component given in Eqs. \ref{eq:rad_velocities} and \ref{eq:rad_velocities1}.
Since coefficient $\alpha$ is taken from $F^{\rm 0}_{\nu}\propto \nu^{\alpha}$, we assumed that $\alpha=3/4$.

We calculate for each component the emitted flux including the Doppler beaming effect as:
\begin{equation}
F^i(\lambda)=F_0^i(\lambda)\{1+[(3-\alpha)(v_i(t)/c)]\},
\label{eq:F_lambda}
\end{equation}
where
\begin{equation}
F_0^i(\lambda)={L^i(\lambda)\over {4\pi D^2}},
\label{eq:F_lambda_1}
\end{equation}
where $D$ is the cosmological luminosity distance to the SMBBH system. In order to avoid the influence of distances, we are going to present our results using luminosity, i.e.,:
\begin{equation}
{4\pi D^2}\cdot F^i(\lambda)=L^i(\lambda)\{1+[(3-\alpha)(v_i(t)/c)]\},
\label{eq:F_lambda_2}
\end{equation}
where $i=1,2$ denotes the component of the system.
The Doppler beaming effect is applied to both - continuum and line emissions.

\subsection{Physical properties of BLRs and broad line parameters}
\label{sec:kinematicsBLR}

The BLR may be present around both SMBH components in the binary system \cite[see discussion in][]{pop12, sp16}. Here we assume that both components have its own BLR, and we consider the following two cases:

\begin{itemize}
\item \textbf{Two separated BLRs}.  There is no contact between BLRs, i.e.,  each BLR is inside the Roche lobe of its SMBBH component (illustrated in Fig. \ref{fig:SMBBH_sketch}). The continuum luminosity of the central source (i.e., accretion disc) determines the dimensions of its BLR (see \S \ref{sec:blr_size}), and the perturbation can be only due to gravitational interaction.
\item \textbf{ Two BLRs inside the Roche lobes of each component and one circum-binary BLR (cBLR)}. Each component has the BLR which is filling the corresponding Roche lobe, and also the total continuum luminosity coming from the accretion disc of both components is high enough to create an extra cBLR (see Fig. \ref{fig:roche_sketch} and \S \ref{sec:BLR_structure}).
\end{itemize}

In both cases we suppose that: i) the BLR is flattened in the same plane as accretion discs, ii) the inner parts of the BLR are overlapping the accretion disc, iii) the BLR extends a few tens of light days in diameter \citep[][]{Kaspi05}, and iv) the kinematics of the BLR depend on the mass of the central SMBH \citep{pet14}. In the second case of separated BLRs, the standard cBLR parameters depend on the dynamical parameters and masses of both components.

As we aim to explore the variability of emission lines usually covered in the optical monitoring campaigns, thus we consider here the H$\beta$ line, which is the most analyzed line in this spectral range.

\subsubsection{Estimation of the BLR sizes}
\label{sec:blr_size}

We use the results of reverberation mapping to constrain the BLR parameters and H$\beta$ line properties, similar as in \cite{sp16}, taking into account the SMBH mass of each component in the binary system, i.e., their emitted continuum luminosities. The reverberation mapping gave the empirical relationships between the BLR size (estimated from the time delay between the continuum and emission line light curves) and the luminosity of the continuum or the broad line \citep[see e.g][and reference therein]{pet14}. Here we use these empirical results to constrain the dimensions of the BLR based on the continuum emission of each component. And then, to estimated BLR dimension, we calculate the intensity of the broad emission line.

There are several empirical relationship between the BLR size and the luminosity of the line and continuum obtained from the reverberation mapping (for review see \cite{pop20}).

For the first case of separated BLRs, one can estimate the H$\beta$ BLR size $R_{\mathrm{BLR}}$ using this relationship given in \citet[][]{Kaspi05}, (see also \cite{sp16}):
\begin{equation}
\frac{R_{\mathrm{BLR}}}{10\textrm{ld}}=
(2.21\pm0.21)\left[\frac{\lambda L_{\lambda}(5100{\textrm{\AA}})}{10^{44}\textrm{erg s}^{-1}}
\right]^{0.69\pm0.05},
\label{eq:rblr}
\end{equation}
where $\lambda L_{\lambda}(5100{\textrm{\AA}})$  is the luminosity at 5100 $\textrm{\AA}$.
Here we use the continuum emission at $\lambda5100$ \AA\ for each of the two accretion discs and apply the above Eq. \ref{eq:rblr}, though there are other relationships \citep[see for a review, e.g.][]{pop20}, which would not change our analysis. Similar computations could be conducted for the high-luminosity/high redshift quasars, using empirical relations presented in e.g., \cite{Kaspi07}.

For the second case, when the binary system is surrounded by additional circum-binary BLR, i.e., cBLR, we use the same Eq. \ref{eq:rblr} for estimating the size of each BLR. In this case of cBLR, the continuum luminosity $L_{\rm total}(5100{\textrm{\AA}})$ is taken to be the sum of luminosities of each disc. For different orbital phases, $L_{\rm total}(5100{\textrm{\AA}})$ will be different, and consequently, the variability in the surrounding BLR should be detected.
This variability is caused by Doppler effect, temperature perturbation of accretion disc and different accretion rates for both SMBHs. Note here, that in difference with circum-binary accretion discs
\citep[see][]{ar94}, the cBLR represents a number of non-interacting emitting clouds which are only gravitationally bound to the central mass (which is sum of masses of two SMBHs) and only follow the gravitational potential of the system, i.e., the inner radius of the cBLR is determined by the Roche lobe.

\subsubsection{The H$\beta$ line profiles - intensity, width and shift}
\label{sec:hb_line}

In general, we can consider different shapes and inclination of the BLR, but one highly expected scenario is that the BLR has a flattened geometry and is strongly self-shielding near the accretion disc \citep[i.e., the AGN equatorial plane, see][]{ga09}. Thus, the BLR is co-planar with the accretion disc, which is also co-planar with the SMBBH orbits (see \S 2.1), and therefore the BLR orientation is defined by the same inclination angle $i$ as the SMBBH system.

The flattened BLR is likely to produce double-peaked line profile, instead of Gaussian profile. However, the most of AGNs show single peaked profiles that indicates that in general BLR is slightly flattened, and that double-peaked profiles are mostly coming from disc-like (strong flattened BLRs). Therefore, here we  assume, for simplicity, that all three BLRs are emitting Gaussian-like line-profiles, whose parameters are reflecting the dynamical effects of the SMBBH system and characteristics of each BLR.

In the case of separated BLRs, i.e., BLRs in Roche lobes of components 1 and 2, we calculate the total line profile of different configurations of the SMBBH system as:
\begin{equation}
I_{\mathrm{dyn}}(\lambda)=I_1(\lambda)+I_2(\lambda),
\label{eq:I_tot}
\end{equation}
where each component emits the Gaussian line profile $I_{i}$ (for $i$=1,2):
\begin{equation}
I_{i}(\lambda)=I_{i}(\lambda_0)\exp{\left[-\left(\frac{\lambda-\lambda_{0}\cdot (1+z^{i}_{\mathrm{dopp}})}{\sqrt{2}\sigma_{i}} \right)^2 \right]}\cos(i)
\label{eq:line}
\end{equation}
where $\lambda_0$ is the transition wavelength for H$\beta$, $\sigma_{i}$ is the velocity dispersion defined by the SMBH mass and dimension of the BLR, and $z^i_{\mathrm{dopp}}$ is the Doppler correction for radial component velocities.

In the case of the cBLR, the total line profile is:
\begin{equation}
I_{\mathrm{tot}}(\lambda)=I_\mathrm{dyn}(\lambda)+I_\mathrm{cBLR}(\lambda),
\label{eq:I_tot1}
\end{equation}
where the cBLR component emits the following Gaussian line profile:
\begin{equation}
I_\mathrm{cBLR}(\lambda)=I_\mathrm{cBLR}(\lambda_0)\exp{\left[-\left(\frac{\lambda-\lambda_{0}}{\sqrt{2}\sigma_\mathrm{cBLR}} \right)^2 \right]}\cos(i)
\label{eq:line1}
\end{equation}
where $\sigma_\mathrm{cBLR}$ is the velocity dispersion defined by the total masses of the SMBBH and dimension of the cBLR.

For calculating the line profile using the above equations, one has to find the maximal line intensity (at $\lambda_0$), velocity dispersion, and line shift. To estimate the  H$\beta$ line intensity we used the empirical relationship given by \cite{Wu04}:
\begin{equation}
\log R_{\mathrm{BLR}}(\textrm{lt-days})=1.381+0.684 \cdot
\log\left(\frac{\lambda L(H\beta)}{10^{42}\textrm{erg\ s}^{-1}}\right),
\label{eq:lumHb}
\end{equation}
from which in combination with Eq. \ref{eq:rblr}, we can derive the connection between the luminosities in the H$\beta$ line and continuum:
\begin{equation}
L_{42}(\mathrm{H}\beta)=C_1\cdot (L_{44})^{C_2},
\label{eq:L_Hb_5100}
\end{equation}
where $L_{42}(\mathrm{H}\beta)$ is the $\lambda L(\mathrm{H}\beta)$ given in units of $10^{42}{\rm erg\ s^{-1}}$, and $L_{44}$ is $\lambda L(5100\mathrm{\AA})$ in units of $10^{44}{\rm erg\ s^{-1}}$. Introduced constants are $C_1=0.88$ and $C_2=1.00877$.

For the Gaussian profile the maximal intensity can be calculated as:
\begin{equation}
I_{i}(\lambda_0)={\lambda L(\mathrm{H}\beta)\over{\sqrt{2\pi}\sigma_i}},
\label{eq:I_lambda_0}
\end{equation}
where velocity dispersion $\sigma_i$ is related to the BLR velocity $v_\mathrm{BLR}$ as:
\begin{equation}
\sigma_i=\lambda_{\mathrm{H}\beta}\frac{v_{\mathrm{BLR}}(m_i)}{c}.
\label{eq:sigma_od_v}
\end{equation}
The velocity in the BLR assuming the BLR virilization can be calculated as:
\begin{equation}
v_{\mathrm{BLR}}(m_i)=\sqrt{\frac{Gm_i}{f_VR_{\mathrm{BLR}}}}.
\label{eq:v_blr}
\end{equation}
where $m_i,\ R_{\mathrm{BLR}}$ and $G$, represents SMBH mass, BLR size and gravitational constant, respectively. $f_V$ is the virialization factor which depends on the BLR geometry and inclination. Here we assumed that whole BLR is virialized, therefore only inclination ($i$) of the BLR is taken into account, then  $f_V=1/\sin^2(i)$ \citep[see][]{af19}. Similarly $v_{\mathrm{cBLR}}$ is calculated taking the total mass ($m_1+m_2$) instead of the mass of one component ($m_i$)
in Eq. \ref{eq:v_blr}.

The gas velocity in the BLR directly depends on the SMBH mass and dynamical parameters.
Thus the line profiles will depend on the same parameters. We underline that the line maximal intensity also depends on the mass of SMBH.

The line shift caused by gravitational effects in the close vicinity of the SMBH is given with \citep[see e.g.][]{jon16}
\begin{equation*}
\Delta\lambda_\mathrm{g}^i=\frac{Gm_i}{cR_{\mathrm{BLR}}}.
\end{equation*}
In comparison to the dynamical shift, the gravitational shift can be neglected for the parameters we use in our model. Therefore we did not take it into account.

\subsubsection{Reduced broad line shift due to Roche lobe}
\label{sec:BLR_structure}

\begin{figure}
\includegraphics[width=9cm]{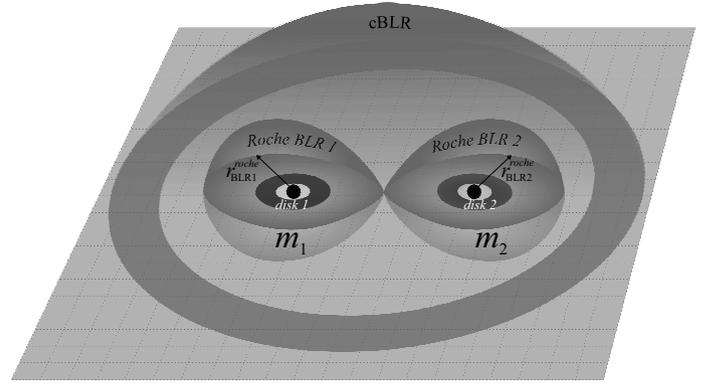}
\caption{SMBBH system in a complex configuration, with the BLRs of each component (BLR1 and BLR2) inside their Roche lobes and the surrounding common circum-binary BLR (cBLR).}
\label{fig:roche_sketch}
\end{figure}

Both components in the assumed binary system can have the BLR with dimensions given by either Eq. \ref{eq:rblr} or Eq. \ref{eq:lumHb}. However, one should take into account that only a part of the BLR which is inside of the Roche lobe of a component will have the radial velocity that reflects the dynamical motion of the component (as it is shown in Fig. \ref{fig:roche_sketch}).
Thus we assume that due to the gravitational pull of an SMBH, only a part of the BLR is dynamically active, i.e., follow the orbital path of the SMBH.

Therefore, the BLR of a single component is reduced to the Roche lobe limit even though empirical relationship gives a larger dimension (see also an illustration in Fig. \ref{fig:SMBBH_sketch}).

The total line profile of a proposed binary system contains three broad
components (see Fig. \ref{fig:roche_sketch}).  The first coming from the
BLR1 - defined by the Roche lobe of the primary component, the second from
the BLR2 - defined by the Roche lobe of secondary, and finally, the third
component  is emitted by the cBLR - the circum-binary BLR surrounding both  
SMBHs (BLR1 and BLR2).  Both SMBHs heat the gas in cBLR, i.e., gas
kinematics is due to the net effect of both masses and luminosities.

This configuration has important effects on the line profiles emitted from BLRs, since the BLR1 and BLR2 are following the dynamics of their SMBHs (orbital motion). Therefore they have radial velocities, while the cBLR is assumed to be a stationary region. Considering the mentioned arguments, in our model, we assume that both SMBHs have their own BLRs  constrained by Roche lobes and are moving with their corresponding SMBH. It is difficult to estimate the exact contribution in the total emission of those moving BLRs since it would require comprehensive magneto-hydrodynamic modeling. Thus we assumed three cases -- the dynamical BLRs contribution is 70\%, 50\%, and 30\%, while the rest of the emission is coming from the stationary cBLR.

\subsubsection{Time delay between continuum and line}
\label{sec:time_delay}

In order to simulate time delay between the variability observed in the continuum and line flux (luminosity), we assumed that the continuum luminosity would affect the dimensions of the BLRs. Taking the continuum time dependent luminosity (as it is calculated using Eq. \ref{eq:SED1}), we account that different luminosity caused different BLR dimensions (see Eq. \ref{eq:rblr}). Therefore $R_{BLR}$ of all three considered regions will change due to changes in the ionization continuum luminosity. We take that the time delay for each BLR is proportional to the light-crossing time over the entire BLR. This can be expressed as:
\begin{equation}
\Delta t_i=R^i_{BLR}/c,
\label{eq:time_delay}
\end{equation}
where $R^i_{BLR}$ is the dimension of i-th BLR and $c$ is the speed of light. With this approximation, we simulate the time lag between continuum and line variabilities. However, this time is significantly shorter than the orbiting period.

\subsection{Algorithm for modeling the composite spectrum of an SMBBH and simulating its variability}
\label{sec:algorithm}

To summarize the proposed model, we list here the steps required to model the composite spectrum of the proposed SMBBH system and simulate its variability:
\begin{enumerate}[(i)]
\item calculate the dynamical parameters and positions of the two components with respect to the barycenter of the SMBBH system (see Eqs. \ref{eq:period1}-\ref{eq:paths});
\item assume inner radius $R_i^{\mathrm{in}}$ and calculate the outer radius $R_i^{\mathrm{out}}$ of both accretion disc (Eq. \ref{eq:rho_out}), which is used to derive the temperature profile across the accretion disc (Eq. \ref{eq:TodR}) and disc luminosity (Eq. \ref{eq:dL}, \ref{eq:SED}), accounting for the relativistic boosting effect (Eq. \ref{eq:F_lambda_2}) and accretion rate (Eq. \ref{eq:TodR});
\item note that the effect of variability is accounted with both the dynamical effects of the binary motion (Eqs. \ref{eq:rad_velocities} and \ref{eq:rad_velocities1}), and the perturbation of the disc temperature (i.e., accretion rate) due to the mutual gravitational interaction of the two components in the SMBBH and change in the accretion rate (Eqs. \ref{eq:Teff_per} and \ref{eq:Teff_pert21} );
\item from the obtained disc luminosity of each component in the SMBBH, we estimate the dimensions of all three BLRs using empirical relationship (Eq. \ref{eq:rblr}), testing whether two separated BLRs are within the Roche lobe of the corresponding central SMBH (\S 2.4.3). If larger, its BLR dimension is reduced to the Roche lobe limit;
\item to simulate the broad H$\beta$ line profile, we assume the Gaussian profile (Eqs. \ref{eq:I_tot}-\ref{eq:line1}) that is based only on the dynamical shift (Eq. \ref{eq:v_blr}). The H$\beta$ line maximal intensity (Eq. \ref{eq:I_lambda_0}) is based on the empirical Eq. \ref{eq:L_Hb_5100}, which is derived from the empirical Radius-Luminosity relationships (Eqs. \ref{eq:rblr} and \ref{eq:lumHb}) and uses the BLR dimensions estimated in the previous step;
\item the composite spectrum (continuum+broad H$\beta$ line) is finally calculated taking into account contribution from all three gaseous regions (BLR1, BLR2, cBLR) with different ratio of the contribution of the surrounding cBLR (70\%, 50\%, and 30\%). Each composite spectrum in one series is normalized to a common constant value, and for this we took the maximal value of the SED of the first spectrum in a series;
\item to explore the spectral variability, we measure the continuum and line luminosity in the H$\beta$ spectral range, which is typically covered in the optical monitoring campaigns, and construct their light curves. After fitting the underlying continuum with the simple power-low from the composite spectrum, the continuum luminosity is measured as the mean value in the narrow interval 5090 -- 5110 \AA\, and the broad H$\beta$ line luminosity only in the range of wavelengths for which the line intensity is greater than zero, when fitted continuum is removed.
\end{enumerate}

To account for the effect of the real measurement uncertainty, we introduce different signal-to-noise (S/N) ratio to the modeled SED, which is later reflected in the measured continuum and line luminosities, i.e., in the continuum and line light curves.
The noise distribution of photons is stochastic in nature and could be modeled with the Gaussian distribution, so-called white noise. Therefore, we added the white noise to the simulated SED (continuum + broad line). Generally, one could use a normal distribution with mean parameter $\mu=0$, and constant dispersion $\sigma_N$, which determines the dispersion of data (additive white Gaussian noise, see \cite{Proakis01}).

Additionally to the white noise, the intrinsic luminosity variability of each component can affect the periodicity signal of the SMBBH. To simulate this effect, so called red noise, we superpose a generated signal to  this noise which is modeled using SER-SAG code \citep[][\href{https://github.com/LSST-sersag/agn_cadences/blob/main/functions.py}{Github link}]{kov21}, which employs Damped Random Walk \cite[DRW,][]{Kelly09}. The
SER-SAG code calculates DRW input parameters a characteristic amplitude $\sigma$, which affects exponentially-decaying variability with time scale $\tau$ around the mean flux $m_{0}$ based on luminosity $L$ of the considered SMBBH component \citep{Kelly09,Kelly13,sub21}. 
The amplitude of the binary signal is taken as a percentage of the mean flux of the considered SMBBH component.

Finally, to illustrate real optical spectra in the H$\beta$ range that are easily distinguished with the prominent doublet of [O III] forbidden lines, we artificially added the narrow [O III] $\lambda\lambda$4959,5007\AA\AA\ lines, calculated from the continuum at 1516\AA\, and using the ratio $\lambda$4959/F$\lambda$5007=1:3 \citep[see][]{dim07}. Note that these narrow [O III] lines were not included in the measurements.

\section{Results and discussion}
\label{sec:results}

To explore the variability of a binary SMBH system which has two accretion discs with two BLRs, and the additional cBLR, we modeled the series of spectra covering the H$\beta$ line emission and the nearby continuum at $\lambda$5100\AA, since this wavelength range is often used in the reverberation mapping and optical monitoring campaigns \citep[see, e.g.][]{ba15,sh16,il17,sh19}. The obtained results are depending only on the dynamics of the system and can be (with small modifications) applied to another wavelength band (as e.g., Mg II broad line and nearby continuum at $\lambda$3000\AA) in case of high-redshift AGNs.

The model is easily adapted to other orientation of the accretion discs and BLRs with respect to the orbital plane of SMBBHs, however, nearly co-planar accretion discs of SMBBHs are expected, especially at sub-pc scales. \citet[][]{sa19} showed for the merger with a high amount of gas, the evolution of the SMBBH system should be connected with their interaction with the surrounding gas. Therefore, the gas accretion onto the SMBBH components results in the alignment of black holes spin with the angular momentum of the binary system \citep[][]{bo07}. Since we consider here sub-pc distances between the components, i.e., final stages of a merger, it is expected that the accretion discs are co-planar. Also, we do not consider extreme mass ratios, thus the timescale after merging are  expected to be long enough that the angular momentum of the accretion discs (and flattened BLRs) aligns with the angular momentum of the circumbinary BLR \citep[similar as in the case of spiraling circumbinary gas, see][]{mi13}. Therefore, for all models presented here, we assumed that discs (and BLRs) are co-planar.

We first present the parameters of the simulated model, then we discuss the variability of the continuum and broad H$\beta$ line, and finally, we give the analysis of the periodicity from the broad line (and different line segments) light curve, which is expected to be detected in the case of the SMBBHs \citep[see e.g.][]{kov18,kov19,kov20}.

\subsection{Parameters of the simulated models}

We performed a number of simulations taking a range of different input parameters of SMBBH systems (different masses, corresponding calculated luminosities and dynamical parameters -- see Table \ref{tab:disc_BLR}). For each set of input configuration, a series of spectra (80 epochs equally spread) is modeled to cover four full orbits of the SMBBH system. As an illustration of the connection between all parameters of an SMBBH system, given here in order to compare dimensions of different regions and orbital parameters, we take the following dynamical parameters: $m_1=1\times 10^8$, $m_2=5\times 10^8$, $e_1=e_2=0.5$, and $a=0.01$ pc (note that this case is not present in Table 1). From these the dimension of accretion disc for components 1 and 2 are: $r_1=4.185\times 10^{-4}$ pc and $r_2$=1.2237$\times 10^{-3}$ pc, respectively. Taking the continuum luminosity at 5100\AA\, from both accretion discs, the BLR dimension for the component 1 is 15 l.d. (0.0134 pc), for the component 2 is 57 l.d. (0.0498 pc), and for the cBLR is 63 l.d. (0.0549 pc). In this case the part of the single BLR covered by the Roche lobe is very close to the accretion discs of components, thus the Roche lobe reduced BLR dimensions (which follow the SMBBH dynamics) are $\mathrm{BLR}1\approx 1 \ \mathrm{l.d.}$ (8.39$\times 10^{-4}$ pc), and $\mathrm{BLR}2= 2.06 \ \mathrm{l.d.}$ ($1.73\times 10^{-3} \mathrm{pc}$),

Table \ref{tab:disc_BLR} lists only the varying parameters for several cases used in our simulations: column (I) gives the order of magnitude of the SMBH mass for both components (in $\mathrm{M\odot}$) and their mass ratio $q$; columns (II) and (III) give the outer radius of the accretion disc of components 1 and 2, respectively; columns (IV) and (V) list the dimensions of the Roche lobe reduced BLR of components 1 and 2 ($\mathrm{BLR}1$ and $\mathrm{BLR}2$), respectively, and column (VI) gives the dimension of the cBLR. All dimensions are given in $10^{-3}$ pc.

\begin{table}
\centering
\caption{Input parameters of the SMBBH systems used in our simulations (column I) and the derived parameters of the dimensions of the accretion discs (columns II and III) and BLRs (columns IV, V and VI), all  given in $10^{-3}$ pc. Given mass $m$ represents the order of magnitude for both SMBHs, expressed in $\mathrm{M\odot}$. The eccentricities are kept constant for all simulations with values $e_1=e_2=0.5$.}
\small
\begin{tabular}{|r|c|c|c|c|c|}
  \hline
  & & & & &  \\[-1ex]
  param. set & $r_1$ & $r_2$ & $\mathrm{BLR}_1$ & $\mathrm{BLR}_2$ & $\mathrm{cBLR}$ \\[2pt]
 (I) & (II) & (III) & (IV) & (V)  & (VI)  \\[2pt]
  \hline
\hline
  & & & & &  \\[-2ex]
  $m\sim10^{6}$ & & & & &  \\
  $q=0.1$ & 0.02 & 0.09 & 0.1 & 0.78 & 0.8  \\
  $q=0.5$ & 0.056 & 0.09 & 0.42 & 0.78 & 1 \\
  $q=1$ & 0.09 & 0.09 & 0.78 & 0.78 & 12  \\[2pt]
\hline
  & & & & &  \\[-2ex]
  $m\sim10^{7}$ & & & & &  \\
  $q=0.1$ & 0.09 & 0.42 & 0.8 & 6 & 6.2  \\
  $q=0.5$ & 0.26 & 0.42 & 3.3 & 6 & 7.7 \\
  $q=1$ & 0.42 & 0.42 & 6 & 6 & 10 \\[2pt]
\hline
  & & & & &  \\[-2ex]
  $m\sim10^{8}$ & & & & &  \\
  $q=0.1$ & 0.42 & 2 & 7 & 43 & 45 \\
  $q=0.5$ & 1.2 & 2 & 27 & 45 & 58 \\
  $q=1$ & 2 & 2 & 47 & 47 & 75 \\[2pt]
\hline
\end{tabular}
\label{tab:disc_BLR}

\end{table}

The case of two separated BLRs in the SMBBH system, and their influence to the line profiles have been investigated in several papers \citep[see, e.g.][etc.]{pop00,sl10,sp16,bog19}, but here we address the long-term variability of the line and continuum emission, which was not studied before and which could be necessary for the future large time-domain surveys, as e.g., LSST. Our simulations show that clearly separated BLRs can influence the line profile, but there is no significant influence to the line and continuum intensity. Therefore here we present the results of our simulations of the sub-pc separated SMBHs in which three BLRs are present.

\subsection{Continuum and broad line variability}

We study the continuum and broad H$\beta$ emission variability and construct their corresponding light curves, and also we explore the mean and \emph{rms} line profiles of broad H$\beta$ line.

\begin{figure*}
\centering
\includegraphics[width=16.cm]{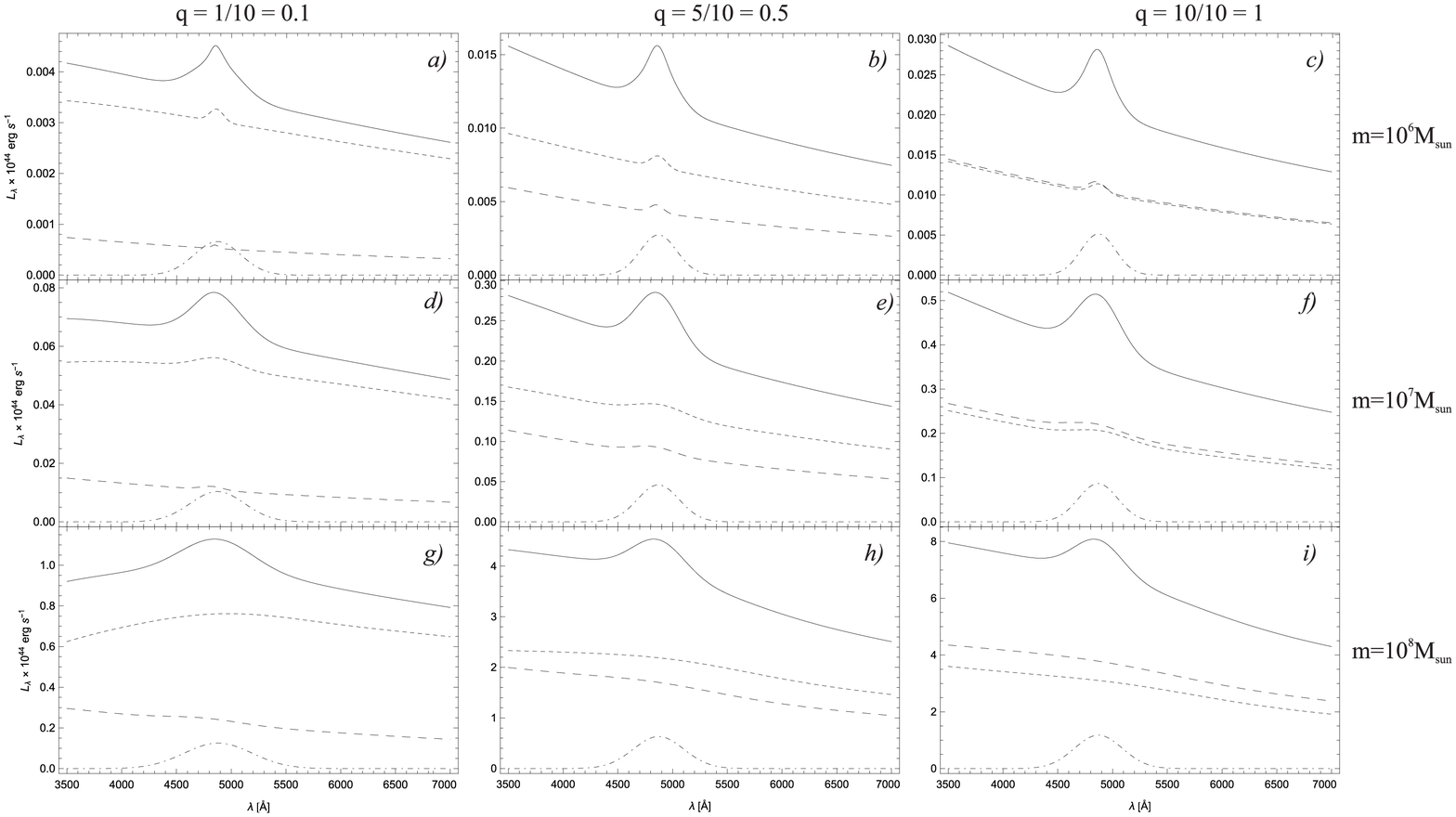}
\caption{Modeled composite spectrum in the H$\beta$ spectral range (continuum+H$\beta$ line) in the closest separation phase ($t=P_{\rm orb}/2$). Different mass ratios
are used $q=0.1,\ 0.5,\ 1$ (from left to right) and different order of magnitude of SMBH masses $10^6,\ 10^7,\ 10^8 \mathrm{M\odot}$ of the secondary component (from top to bottom). The notation is as follows: dot-dashed line - emission from cBLR, long-dashed line - emission from the BLR1 of the component 1, short-dashed line -  emission from the BLR2 of the component 2, solid line - total emission, i.e., the sum of both components emission (line and continuum) and cBLR. Only broad H$\beta$ line is considered (without [O III], He II, narrow H$\beta$, etc.). In this case, 70\% of total BLR emission originating from cBLR.}
\label{fig:Hb_678}
\end{figure*}

\begin{figure*}
\centering
\includegraphics[width=16.0cm]{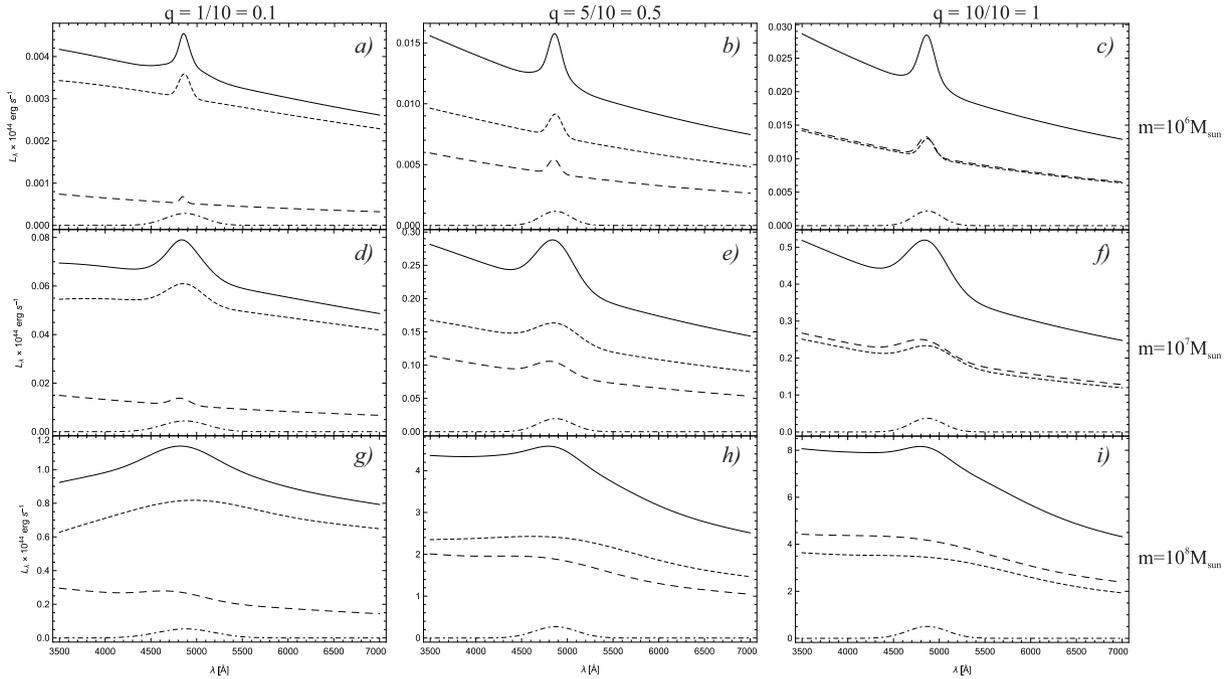}
\caption{The same as in Fig. \ref{fig:Hb_678}, but if contribution of the cBLR to the total emission is 30\%.}
\label{fig:Hb_678-1}
\end{figure*}

The shape of the SED in the H$\beta$ wavelength range is plotted in Fig. \ref{fig:Hb_678}, for different mass ratio ($q=0.1,\ 0.5,\ 1$) and different order of magnitude of SMBH mass ($10^6,\ 10^7,\ 10^8 \mathrm{M\odot}$) of the secondary component. This model accounts for three BLR, two BLRs of primary and secondary components which are coming from the corresponding Roche lobes (BLR1 and BLR2, denoted with the long-dashed and short-dashed line, respectively), and one cBLR component (dot-dashed line in Fig. \ref{fig:Hb_678}). Fig. \ref{fig:Hb_678} shows cases in which 30\% of the line luminosity is coming from the Roche lobe emission and 70\% from the cBLR. In Fig. \ref{fig:Hb_678-1} we show the opposite case, i.e., the contribution from the cBLR is only 30\%.

Figs. \ref{fig:Hb_678} and \ref{fig:Hb_678-1} show no significant difference in total broad line profiles between cases with different cBLR and Roche BLRs (BLR1, BLR2) contributions. However, the broad line profiles, as well as the continuum and line intensities are  significantly changing not only with the mass ratio, but also with the black hole mass of the components. However, in the case of the massive components (order 10$^8\mathrm{M\odot}$), the BLR components coming from the Roche lobes for subpc distant SMBBHs are contributing to the continuum around the H$\beta$ line, while for smaller masses, Roche components of the BLR are contributing to the H$\beta$ line. This is very important, since, the expected dynamical effects in the broad  line profiles probably are not present in the case of subpc massive SMBBHs ($\sim 10^8\mathrm{M\odot}$). Also, it seems that only single cBLR is significantly contributing to the observed broad line profiles (see plots {\it i)} in Figs. \ref{fig:Hb_678} and \ref{fig:Hb_678-1}).

\begin{figure*}
\centering
\includegraphics[width=17cm]{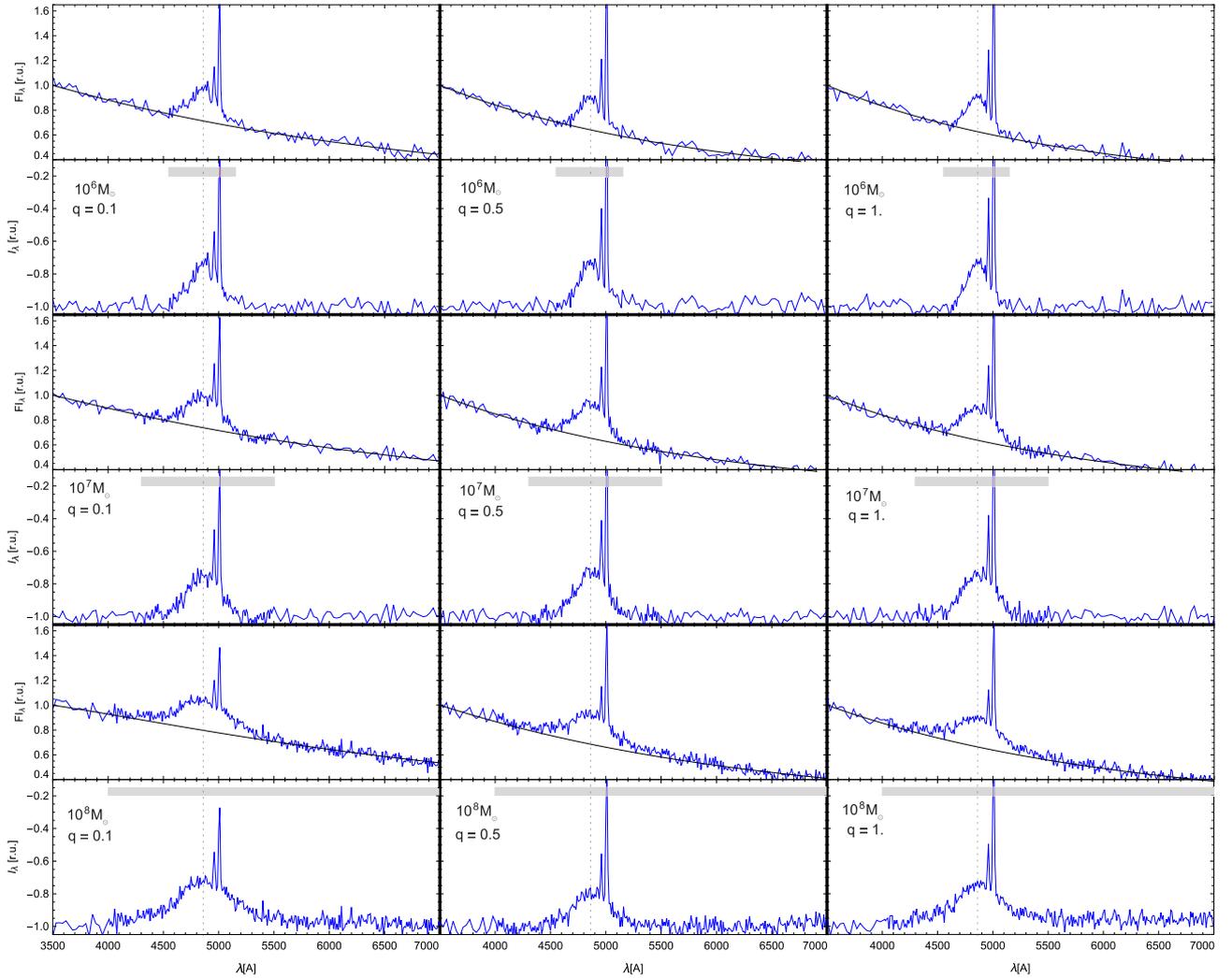}
\caption{The same as in Fig. \ref{fig:Hb_678},
but including the effect of noise (3\%) and [O III]$\lambda\lambda$4959,5007\AA\AA\
emission in order to illustrate the real AGN spectra typically observed in the optical band. For each case of input parameters (indicated on the panels) two sub-panels are shown, with the modeled composite spectrum (upper) and continuum subtracted (bottom). The cBLR contributes with 30\% to the total emission and perihelion position of SMBBH is assumed.} Shaded area defines the limit of the H$\beta $line integration, and vertical line indicates the position of H$\beta$.
\label{fig:lc_678_01_05_1_per}
\end{figure*}

\begin{figure*}
\centering
\includegraphics[width=17cm]{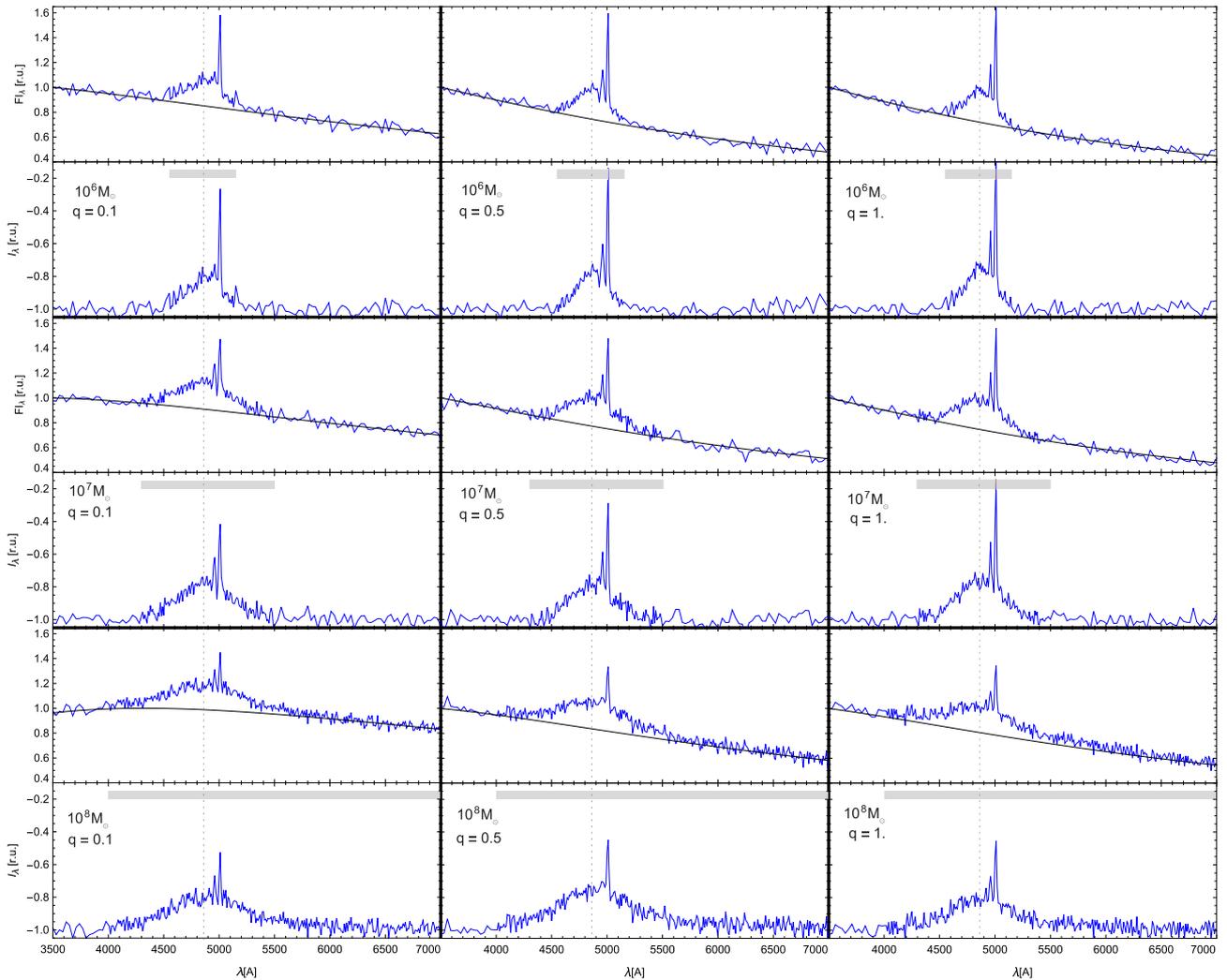}
\caption{The same as in Fig. \ref{fig:lc_678_01_05_1_per}, but for aphelion position.}
\label{fig:lc_678_01_05_1_aph}
\end{figure*}

The simulated spectra with the added white noise of 3\% ($\sigma_N=0.03$) and artificially added [O III] $\lambda\lambda$4959,5007\AA\AA\ emission for different masses and mass ratio is present in Figs. \ref{fig:lc_678_01_05_1_per} and \ref{fig:lc_678_01_05_1_aph}. We considered broad line profiles for two positions of components, at minimal and maximal distances. For each case of input parameters two sub-panels are shown, with the modeled composite spectrum (upper sub-panels) and continuum subtracted (bottom sub-panels). There is a difference in line profiles, in the case of minimal distances, the line profiles are weaker and narrower (Fig. \ref{fig:lc_678_01_05_1_aph}), compared to the maximal distance of the SMBHs, when the rate of accretion produces more intensive and broader lines (see Fig. \ref{fig:lc_678_01_05_1_per}). This cannot be expected in the case of an AGN with a single SMBH, for which  the line luminosity is connected with the size of the BLR , i.e., when the broad line is weak,  the BLR radius decreases, the emission region is closer to the central SMBH that causes higher rotation and as a result, the line width will increase. However, in close binary SMBH systems with  complex BLR (composed of three BLRs), the situation is the opposite. The minimal distance between the binary components, lines emitted by the BLR1 and BLR2 are very broad and shifted (due to high radial velocities). This causes that these line components contribute mainly to the continuum and marginally to the total line flux, whereas in the case of the maximal distance, when accretion is the highest, the radial velocities of BLR1 and BLR2 components are smaller and their contribution to the total line is important (especially to the line wings). This effect produces a more intensive and broader line in the case of maximal component distance than in the case of minimal one. 

Also, Figs. \ref{fig:lc_678_01_05_1_per} and \ref{fig:lc_678_01_05_1_aph} demonstrate that the added noise influences the modeled spectrum. But, if the noise is on the level of several percent, the influence is not so strong, especially in the line profile, which is still clearly detected. However, the spectrum is affected by different S/N, and this especially manifests in the measured continuum, line parameters, and their variability (see next sections). 

\begin{figure}
\includegraphics[width=8.5cm]{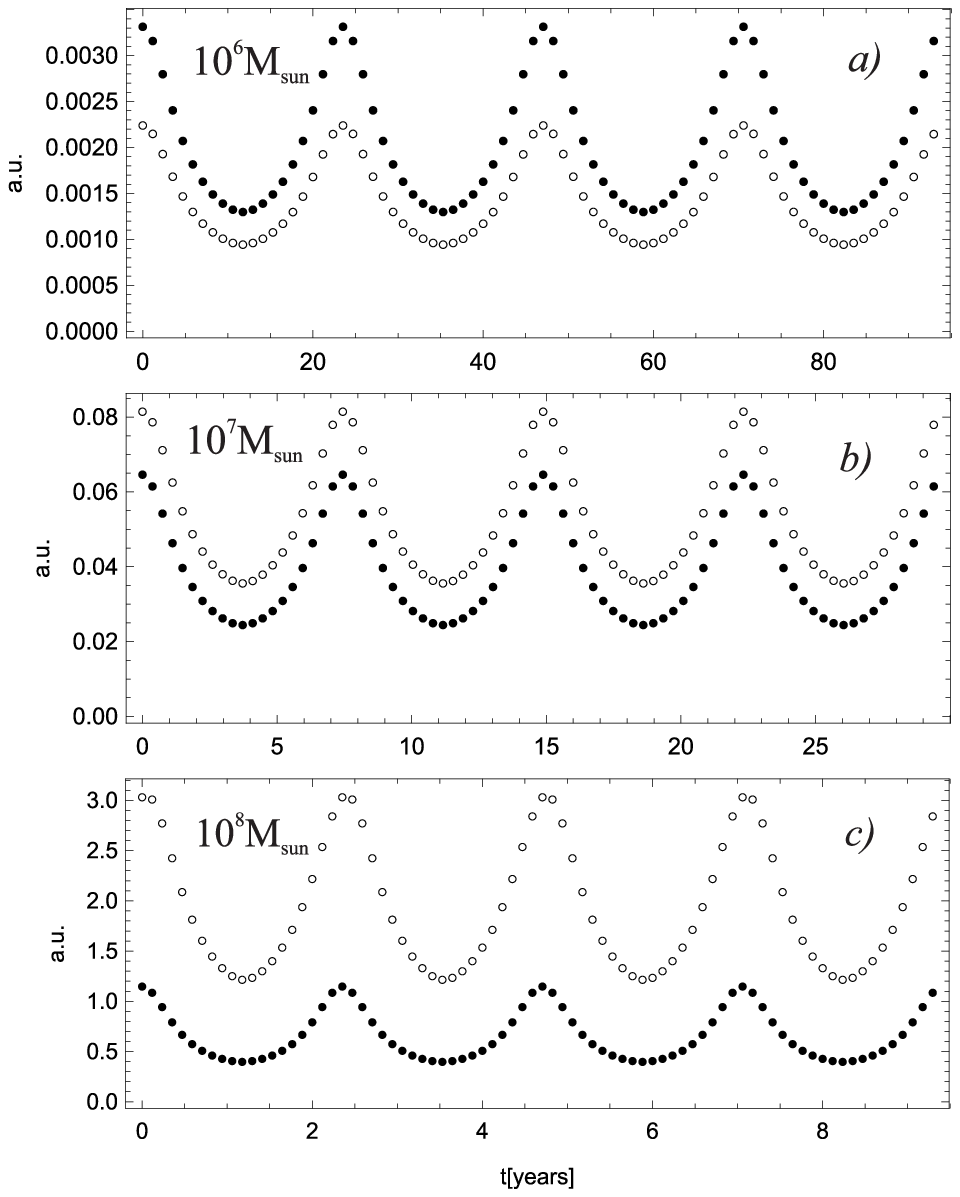}
\caption{Luminosity in  arbitrary  units of the total H$\beta$ line (open circles) and continuum luminosity (full circles), for different order of magnitude of SMBH mass of the more massive component ($10^6\mathrm{M\odot}$, $10^7\mathrm{M\odot}$ and $10^8\mathrm{M\odot}$, from top to bottom) and mass ratio $q=0.5$. Intensity is scaled to fixed value for all cases. }
\label{fig:lumHbcont}
\end{figure}

\subsubsection{Continuum and line luminosity light curves}

We modeled broad H$\beta$ line and continuum light curves of the SMBBH system, taking four full rotation phases, for 80 equally distributed epochs. These simulated light curves are presented in Fig. \ref{fig:lumHbcont}, which shows broad line (open circles) and continuum (full circles) light curves for the mass of more massive component $m_2$ of the order of $10^6\mathrm{M\odot}$, $10^7\mathrm{M\odot}$ and $10^8\mathrm{M\odot}$ (from top to bottom), with the component mass ratio $q=0.5$ and cBLR contribution of 30\%. Line and continuum luminosities are normalized to the maximal value during the simulated period.

The line variability, which is caused by dynamical effects, is more prominent than the continuum one (Fig. \ref{fig:lumHbcont}). 
The line and continuum variability increase with total component mass and the periodicity can be detected in both, line and continuum light curves. 
However, the line variability in the massive SMBBHs seems to be more prominent than the continuum one.

In all three cases the variation of the H$\beta$ line is very high and show similar trends (expressed in arbitrary units), therefore, one could expect that the periodical variability in broad lines is more perspective for detection than from the continuum light curves.

\begin{figure}
\centering
\includegraphics[width=8.5cm]{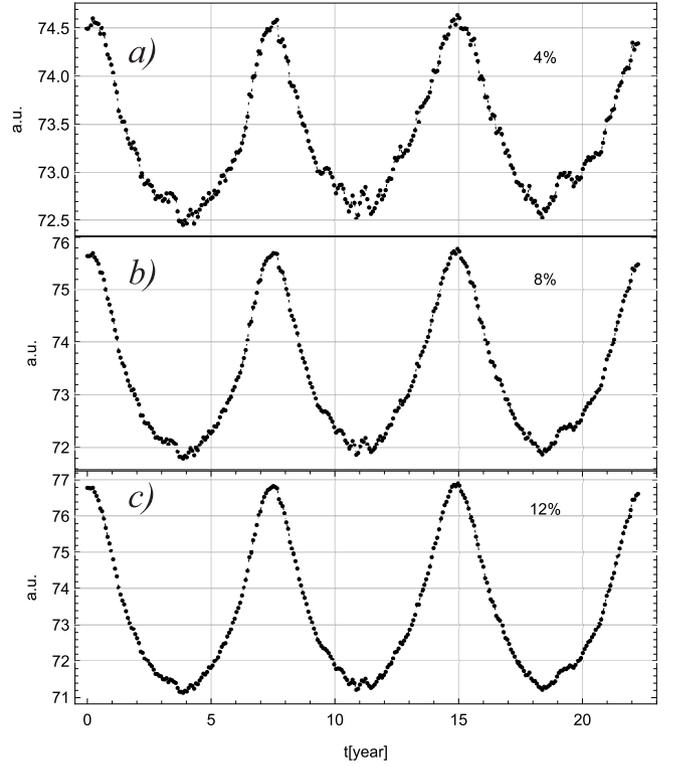}
\caption{Continuum flux at $\lambda=5100${\AA} (case \emph{b)} in Fig. \ref{fig:lumHbcont}), for three different cases of signal strength (4, 8 and 12\%) in comparison to the superposed red noise.}
\label{fig:cntRN}
\end{figure}

We generated periodic signal from our model and added red noise based on Eqs. 1 and 13, so that the final light curve is: $F(t)=drw+A\cdot F_{\rm tot}$, where A is modulation amplitude, $drw$ is red noise, and $F_{\rm tot}$ is binary periodic signal. 
We make a computations for case \emph{b)} in Fig. \ref{fig:lumHbcont}  and present it in Fig. \ref{fig:cntRN} using arbitrary units. 
Amplitude is taken to be $A=perc \cdot \mathrm{Mean}(drw)$, where $perc = 4, 8,  12\%$ (see Fig. \ref{fig:cntRN}).
We can see that decreasing the amplitude of periodic signal the red noise becomes dominating and vice versa.

\subsubsection{Broad line profile variability}

\begin{figure*}
\centering
\includegraphics[width=17cm]{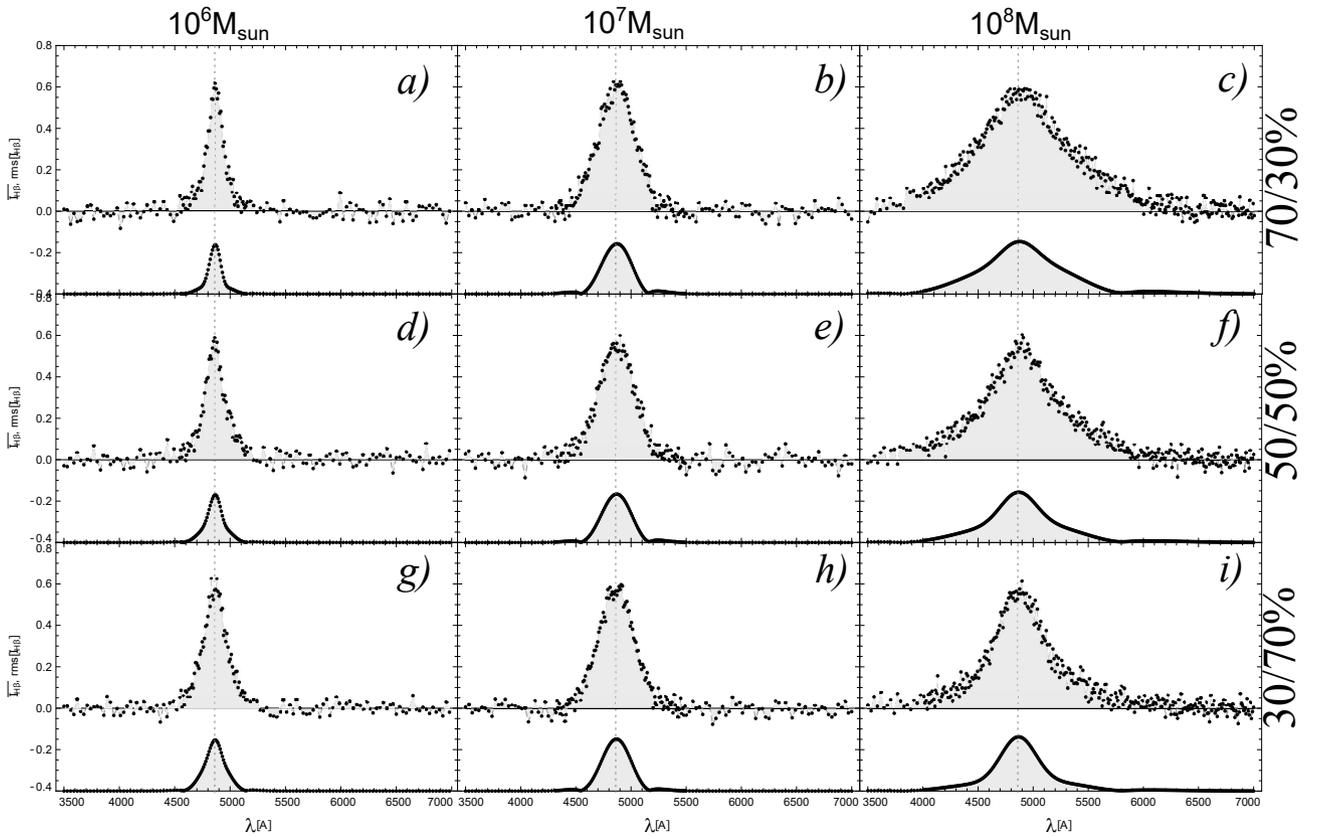}
\caption{Mean profile of H$\beta$ line during the four full rotation of the system, for different order of magnitude of SMBH masses ($10^6\mathrm{M}_{\odot}$ to $10^8 \mathrm{M}_{\odot}$ from left to right), and different ratio of contribution to the total line luminosity from the BLRs of both components (BLR1, BLR2) and cBLR (70\% coming from the BLR1+BLR2 and 30\% from the cBLR, 50/50\%, and 30/70\% from up to bottom). The \emph{rms} profile is given at bottom of each panel. We remind the reader that the orientation of the system (orbital inclination $i=45^{\circ}$) is always the same}.
\label{fig:Roche_Hb_05}
\end{figure*}

For the analysis of the variability in the line profiles, the monitoring campaigns usually use the mean and \emph{rms} profiles, which reflect the changes in the line emitting region.
We calculate the H$\beta$ line profile for different binary phases for the same parameters as in previous analysis (three different masses $10^6,\ 10^7,\ 10^8\ \mathrm{M}_{\odot}$ and the same mass ratio $q=0.5$).
After that we construct the mean and corresponding \emph{rms} line profile, which are shown in Fig. \ref{fig:Roche_Hb_05}. The simulations are performed for three different masses and three different ratios of contribution to the total line luminosity of the BLRs of both components (BLR1 and BLR2) and additional cBLR (ratios are denoted on the right axis). In this computation we superposed 3\% of additional white noise on artificially generated spectrum. Since in this analyze, attention is on the line profile, we normalize flux to maximal value of continuum for each particular case.

Fig. \ref{fig:Roche_Hb_05} shows that mean line profiles change according to mass ratio and ratio of emission from BLRs. For higher component masses, the line profile is broader than in lower mass case. Additionally, line profile deviates from a Gaussian shape and asymmetry of line is more prominent in the case of higher component masses. For example, an order of $10^8\mathrm{M_{\odot}}$ component masses, the emission coming from Roche BLR (BLR1, BLR2) contributes to the wings, whereas the cBLR emission contributes to the line core. This is in accordance with the dynamics of the BLRs, since Roche BLRs are moving and therefore suffer Doppler effect, while cBLR is stationary.

However, the \emph{rms} profiles show that most variability is not in the line wings, but in the line core (Fig. \ref{fig:Roche_Hb_05}, bottom of each panel). This is because most of the emission in the line core comes from an extended cBLR that is excited by continuum emission from both accretion discs. 
Additionally, the \emph{rms} does not follow Gaussian shape particularly in cases with Roche BLRs domination. 

\begin{figure*}
\centering
\includegraphics[width=16cm]{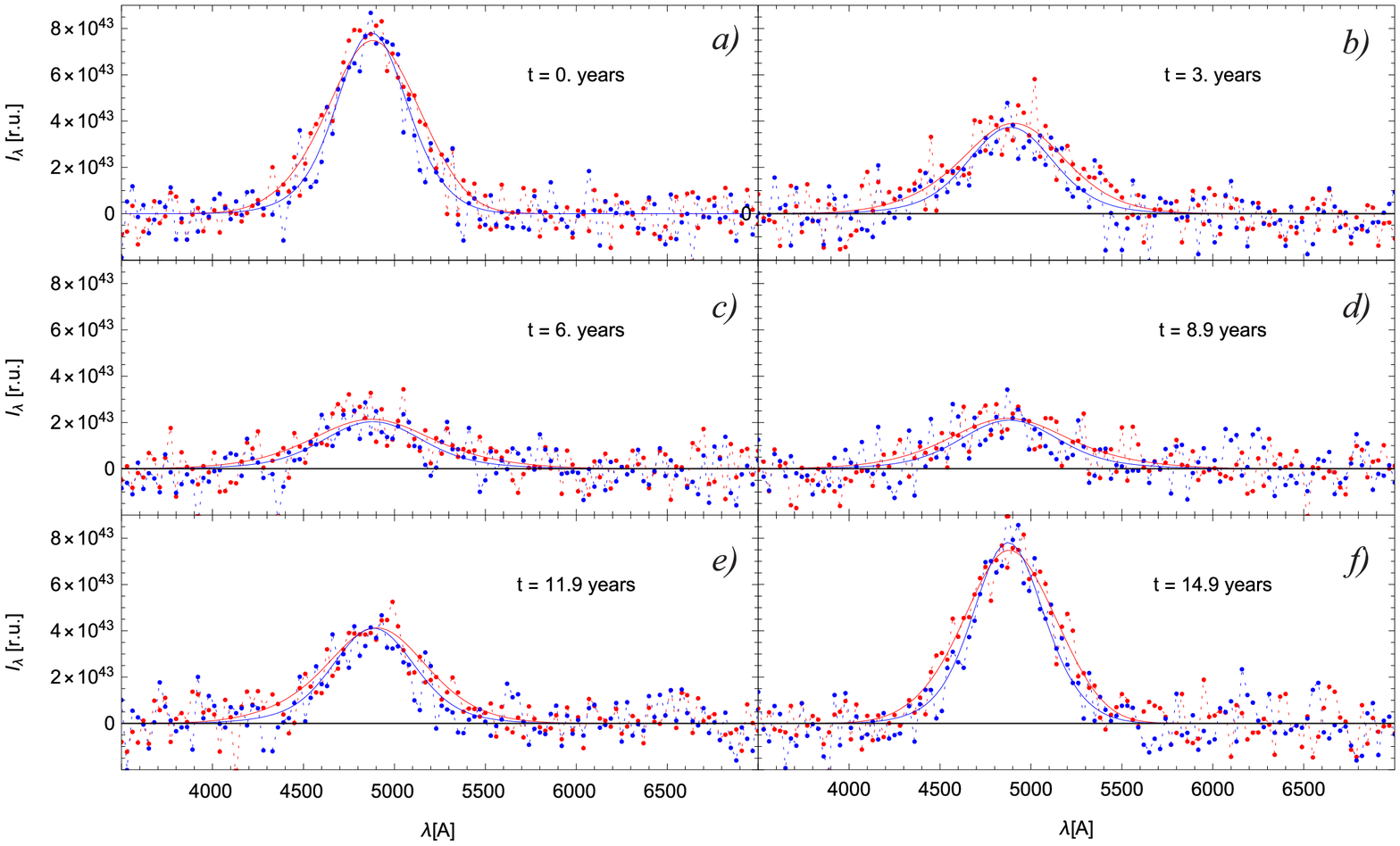}
\caption{H$\beta$ spectral line in six different phases of the SMBBH system with following parameters: $m_1=1\times10^{8}\mathrm{M\odot}$, $m_2=5\times10^{8}\mathrm{M\odot}$, $a=0.01$pc and $e=0.5$, with included white noise ($\mu=0$, $\sigma=0.03$). Blue points indicate the case when the cBLR contributes with 70\% to the total emission, and the red points with 30\%. Blue and red lines designate H$\beta$ spectral line flux, without white noise included. Color coding is same as dots designation.}
\label{fig:Hb_phase}
\end{figure*}

\begin{figure}
\centering
\includegraphics[width=8cm]{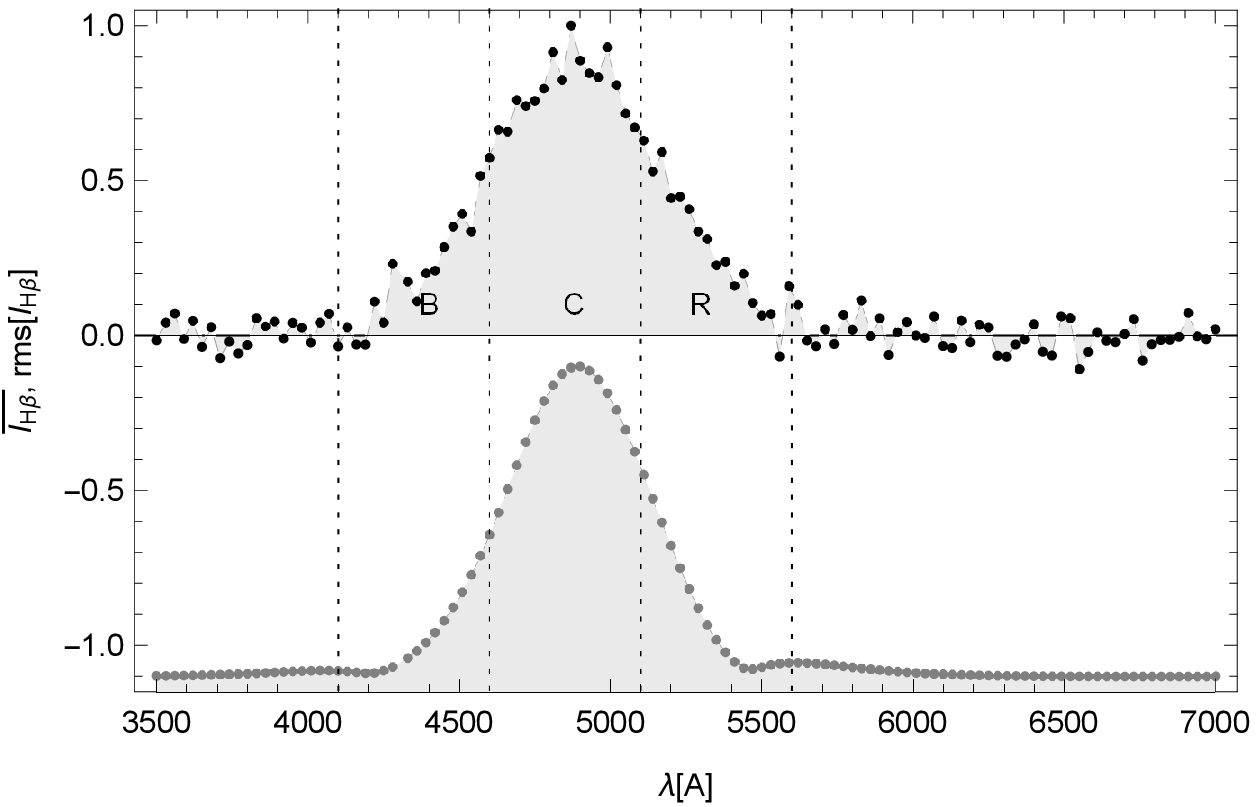}
\caption{Mean profile (upper panel) and \emph{rms} value (lower panel) of six H$\beta$ line profiles (case when the cBLR contributes with 30\%) presented in Fig. \ref{fig:Hb_phase}.
Dotted lines divide the mean profile in three different segments used in our computations, blue parts (B=$\lambda\lambda$4100-4600\AA), central (C=$\lambda\lambda$4600-5100\AA) and red (R=$\lambda\lambda$5100-5600\AA) segment.}
\label{fig:Hb_mean_sek}
\end{figure}

White noise could also affect the spectral line profiles. Since, the shape of the broad line can indicate an SMBBH system, here we explore the
influence of white noise to the line profile of H$\beta$
(see  Fig. \ref{fig:Hb_phase}). We add the white noise on H$\beta$ using $\mu=0$ and $\sigma=0.03$. 
The specific line profiles for six different epochs are shown in Fig. \ref{fig:Hb_phase}. The line profiles and intensities are affected by the noise, but still one can clearly recognize the different line profiles in different phases (Fig. \ref{fig:Hb_phase}), which are the product of different positions and dynamical parameters of the components. The mean and \emph{rms} line profiles constructed from the
six characteristic orbital phases shown in Fig. \ref{fig:Hb_phase} are given in Fig. \ref{fig:Hb_mean_sek}. The mean line profile shows nearly symmetrical shape, with approximately Gaussian profile, with slight deviation in line wings. The \emph{rms} profile shows a maximum at the line center indicating maximal variability, contrary to the line wings where variability is significantly lower. 

The mean H$\beta$ line profile is a sum of three components, two which are emitted from the BLRs that surround the SMBHs (within each Roche lobe) and one from the cBLR that is enveloping the whole binary system. Since Roche lobe follows the dynamics of SMBHs, these components could have high radial velocities and consequently a significant contribution from relativistic boosting effect. As a result for higher mass systems a significant change may be present in the  line wings. However, since the change in the accretion is the most important, and it has maximum  at maximal distances between component, the amplification of the cBLR is more prominent than in Roche BLRs, therefore the variability in the line core is dominant. 

\subsubsection{The line vs. line segment variability}

In order to investigate the flux variability across the line profile and its correlation with the continuum, we divide the H$\beta$ line in three segments (case when the cBLR contributes with 30\%): blue B=$\lambda\lambda$4100-4600\AA, central C=$\lambda\lambda$4600-5100\AA and red segment R=$\lambda\lambda$5100-5600\AA , as it is shown in Fig. \ref{fig:Hb_mean_sek}. The continuum emission at $\lambda=5100$\AA, is used in this analysis.

\begin{figure}
\centering
\includegraphics[width=8cm]{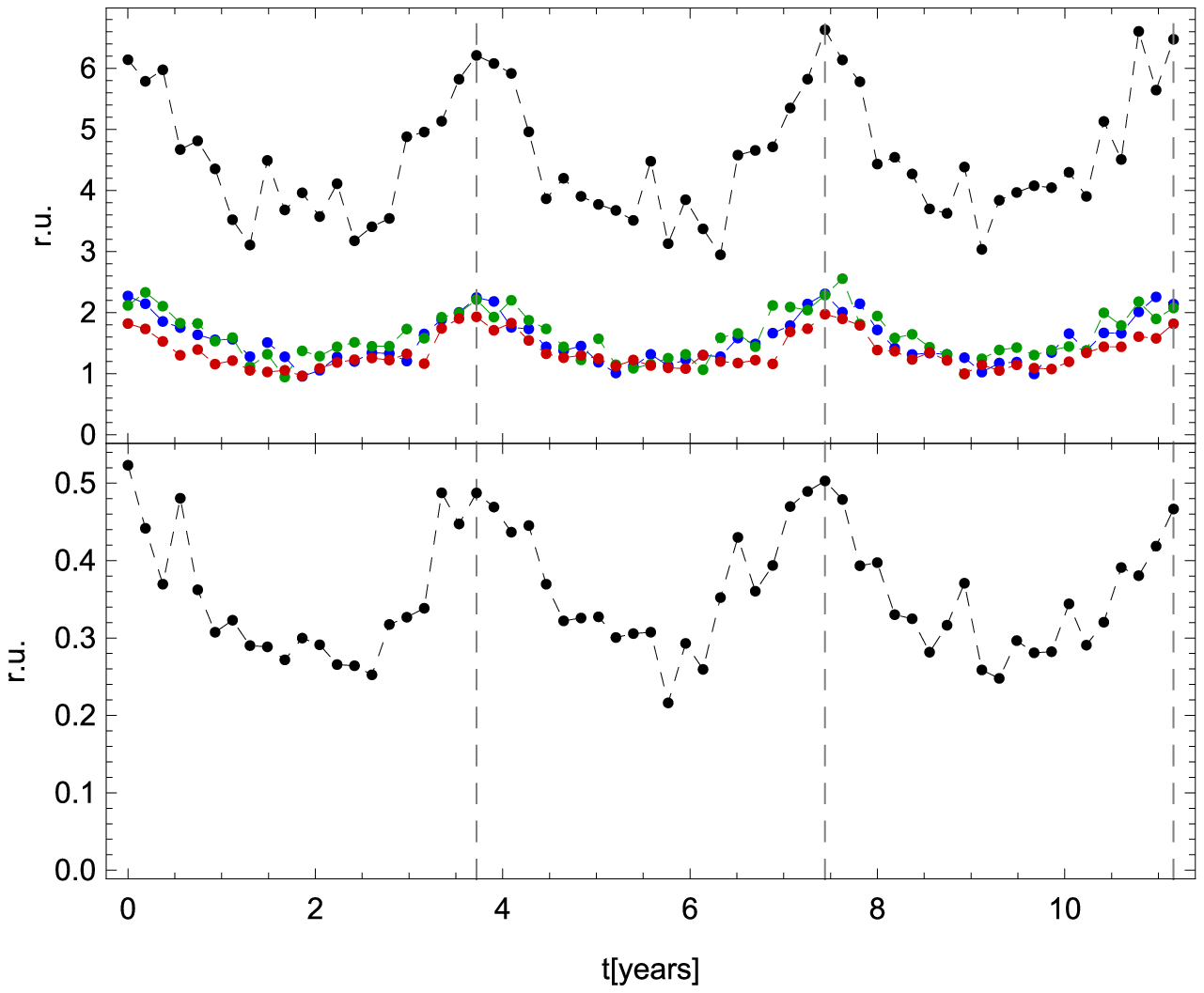}
\caption{Light curves of different line segments, blue (blue dots), central (green dots), red (red dots) segments and total (black dots) H$\beta$ line emission (upper panel), and continuum at $\lambda5100$\AA\ (bottom panel). The white noise on the level of 3\% is taken into account,  and the red noise is applied as described in section \ref{sec:algorithm}}. 
\label{fig:rel_var_bcr_cont}
\end{figure}

The light curves (Fig. \ref{fig:rel_var_bcr_cont}) are simulated taking three full orbits of the system with following parameters $m_1=1\times 10^7\mathrm{M\odot}$ and $m_2=5\times 10^7\mathrm{M\odot}$, eccentricity
$e_1=e_2=0.5$, and the mean distance of $a=0.01$pc. We add white noise with parameters $\mu=0$ and $\sigma=0.03$ to the simulated light curves. Furthermore, we assumed red noise influence on light curve signals as  described in section \ref{sec:algorithm}. The total flux is in arbitrary units and it is given in absolute scale. 
Fig. \ref{fig:rel_var_bcr_cont} reveals that light curves of line segments vary in a similar way.
The periodicity is present in all segments, and in the total line flux (upper panel) as well as in the variation of the continuum (bottom panel).

\begin{figure*}
\centering
\includegraphics[width=17cm]{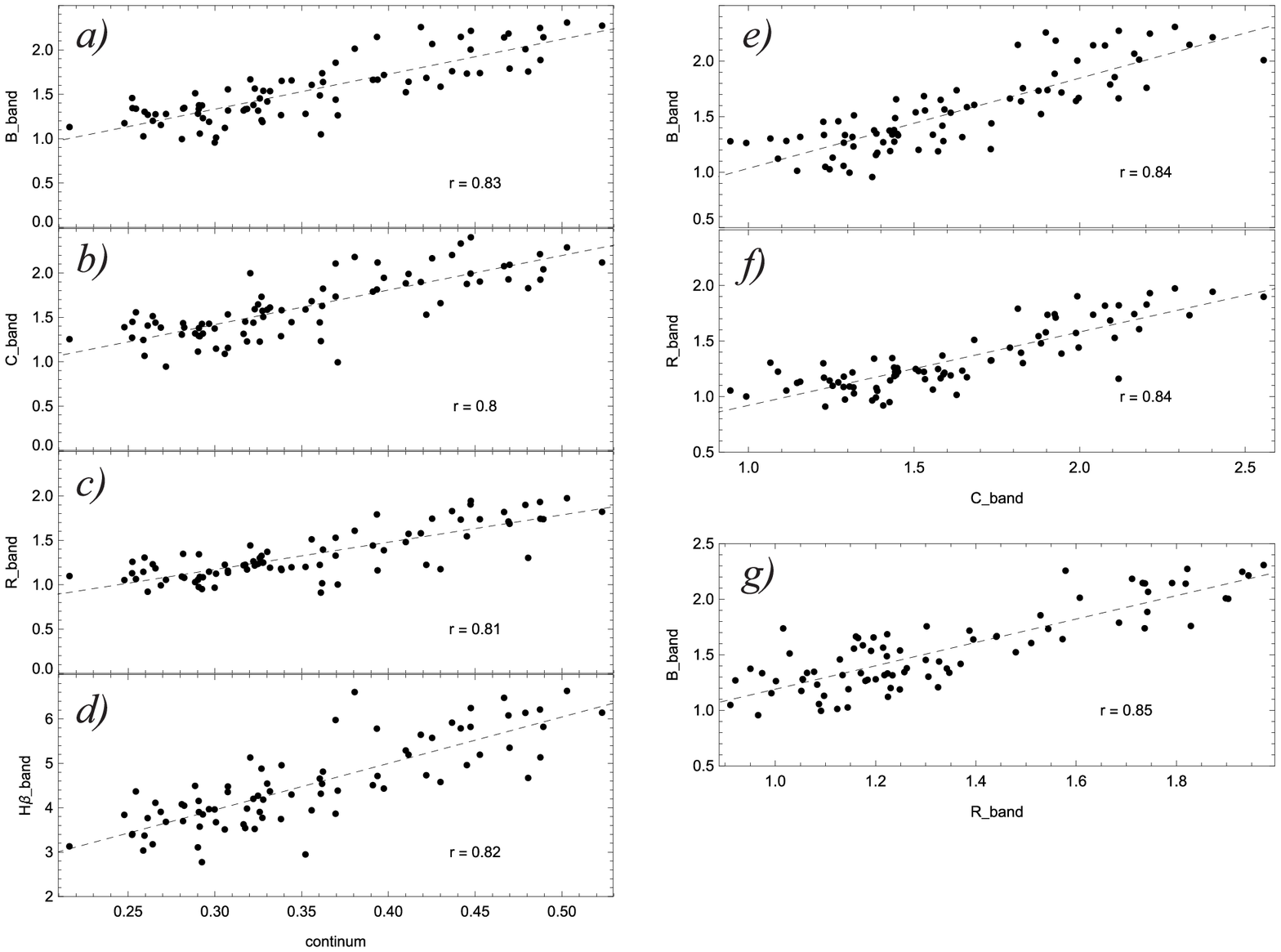}
\caption{Left: correlation between different line segments of H$\beta$ line and continuum luminosity: a) blue, b) central, c) red, and d) total H$\beta$ line. Right: correlation between different segments: blue and red vs. the central line segment (panels e) and f), respectfully). Panel g) presents blue vs. red segment correlations.}
\label{fig:corr_bcrHbcont}
\end{figure*}

We also calculated correlations (based on the Pearson's correlation coefficients) between the H$\beta$ line segments and continuum at $\lambda=5100$\AA. Fig. \ref{fig:corr_bcrHbcont} (panels \emph{a), b), c) and d)}) shows that there is a significant correlation between fluxes in the H$\beta$ line segments (and total H$\beta$ emission) with the continuum, with very similar correlation coefficient of $r\approx 0.8$. 
Correlations between line segments (Fig. \ref{fig:corr_bcrHbcont}, panels \emph{e}), \emph{f}) and \emph{g})) are also very high.

\begin{figure}
\centering
\includegraphics[width=8.5cm]{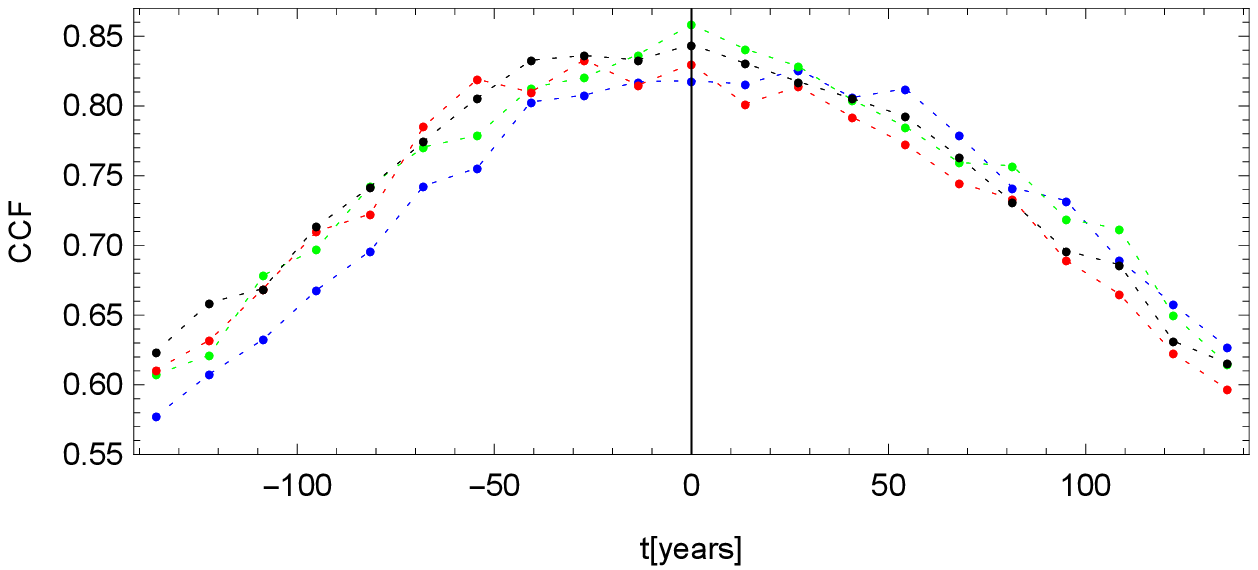}
\caption{Cross correlation function between blue (blue dots), central (green dots), red (red dots), total (black dots) line segments and continuum at $\lambda=5100\mathrm{\AA}$.}
\label{fig:ccf}
\end{figure}

Additionally, to confirm these correlations we computed a cross correlation function between H$\beta$ line segments and continuum, shown in Fig. \ref{fig:ccf}. The high correlation between line segments (blue, green, red and total H$\beta$) and continuum is seen. 
It is noticeable that there is a slight time delay of the blue segment to
the central and red segments. This is expected since we have assumed time delay
between different BLRs and continuum (see \S \ref{sec:time_delay}), reflecting the
difference in the contribution of different BLRs to the emission line parts.
However, the time delay ($\sim$ days) compared to the length of the light curve
($\sim$ years) is too short and cannot be easily detected by simple visual inspection
of the continuum and line light curves. To carefully examine time delays, one
should model the light curves with a denser time sampling, which will be done
in the forthcoming paper.

\subsection{Periodicity}

One of the expected observational characteristics of an SMBBH system is the periodicity or quasi-periodicity in the line and continuum light curves during several orbiting periods \citep[see also][etc]{bon12,gr15,gr15a,li16,kov19,kov20}.
However, if the periodicity is detected in an observed light curve, it may indicate the presence of binary system \citep[see][]{kov19,kov20}.

\begin{figure}
\centering
\includegraphics[width=8.5cm]{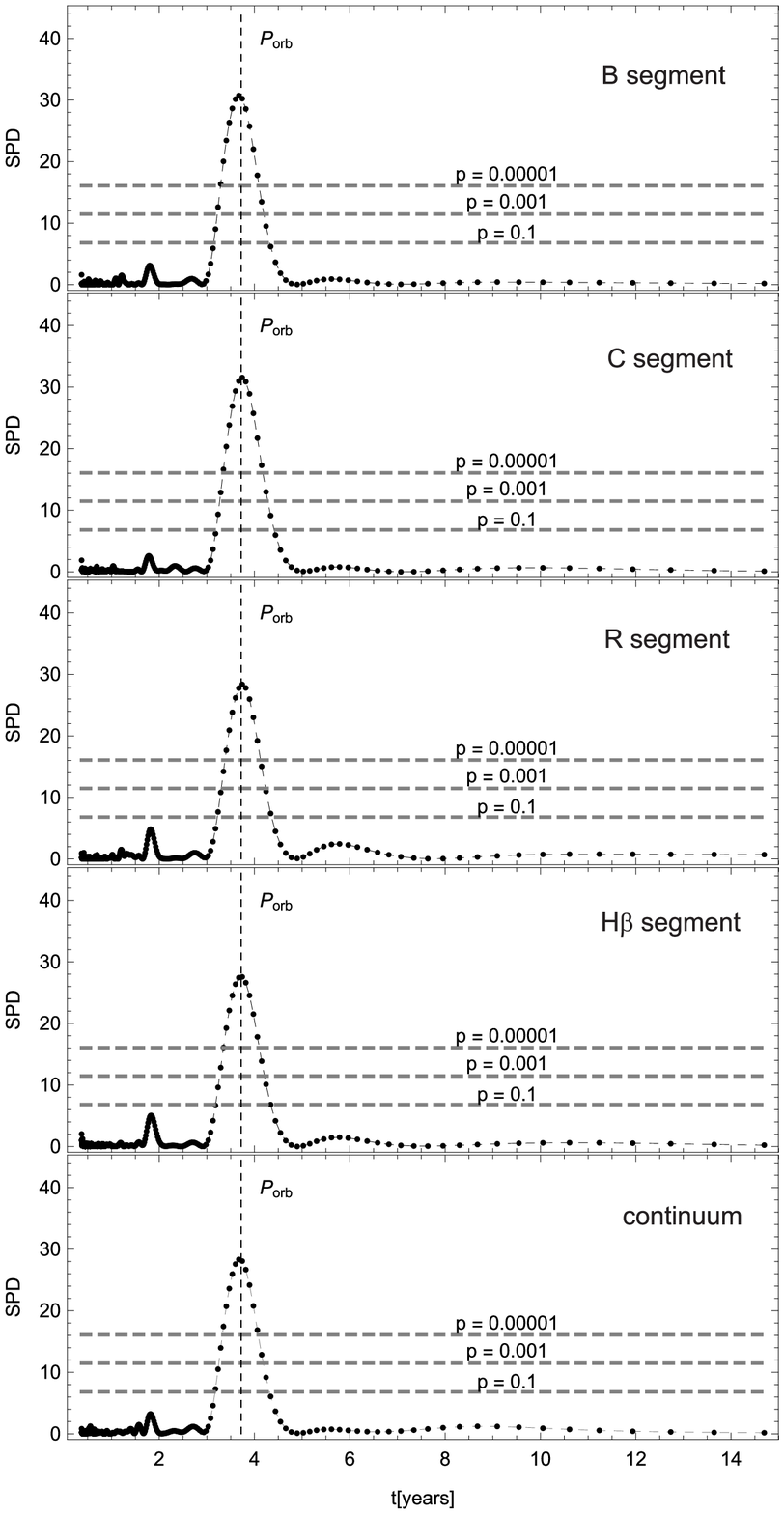}
\caption{Periodograms computed for different line segments (blue, center and red), total H$\beta$ line, and continuum at $\lambda=5100\mathrm{\AA}$.
Vertical dashed lines show the position of $P_{orb}$. Horizontal dotted lines give significance in spectral power density.}
\label{fig:per_bcrHbcont_all}
\end{figure}

Here we perform an initial period analysis using Lomb-Scargle \citep[hereafter LS][]{lo76,sc82} algorithm and compute periodicity diagrams (periodograms) of the continuum and line light curves given in Figs. \ref{fig:cntRN} and \ref{fig:Hb_phase}. More detailed investigations including other methods for periodicity detection, are left for future studies.
In Fig. \ref{fig:per_bcrHbcont_all} we present periodograms for continuum at $\lambda=5100${\AA}, total $\mathrm{H\beta}$ line and its line segments.
Peaks of the periodogram curves show the orbital period of the SMBBH  system. We also plot the horizontal significance level lines for three different values of parameter $p$ (False Discovery Rate), $p=0.1$, $p=0.01$ and $p=0.001$. With higher SPD (Spectral Power Density) values of the peak, false discovery rate decreases. If height of the peak is reaching the value of parameter $p=0.1$, than it means that there is a 10\% probability of mistake. For lower values of SPD, probability of false discovery is bigger, and for higher SPD, false alarm has less probability.

Fig. \ref{fig:per_bcrHbcont_all} illustrates that periodicity could be determined by using any of the line segments or total line, and continuum flux, since there is not much difference in periodicity determination. A major issue here is difference in intensity of the observed flux, since line flux is much stronger than continuum (see Fig. \ref{fig:rel_var_bcr_cont}). 
We outline that the LS peaks are very broad and have deformed levels, due to which it is not possible to determine their positions with high accuracy.

\begin{figure*}
\centering
\includegraphics[width=17cm]{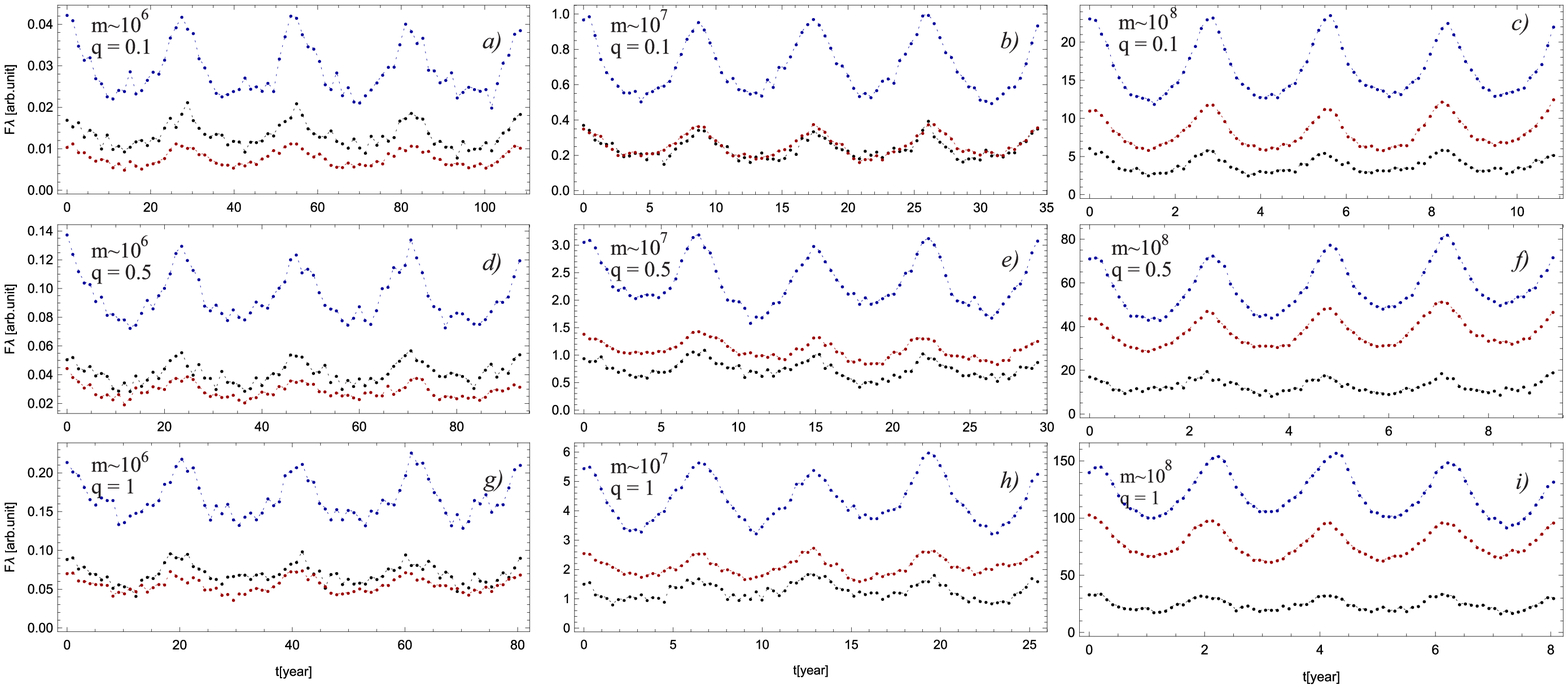}
\caption{H$\beta$ (red with 30\% and blue with 70\% contribution of the cBLR) and continuum (black dots) light curves for the SMBBH systems plotted in the Fig. \ref{fig:lc_678_01_05_1_per}.}
\label{fig:var_678_01_05_1}
\end{figure*}

\begin{figure*}
\centering
\includegraphics[width=17cm]{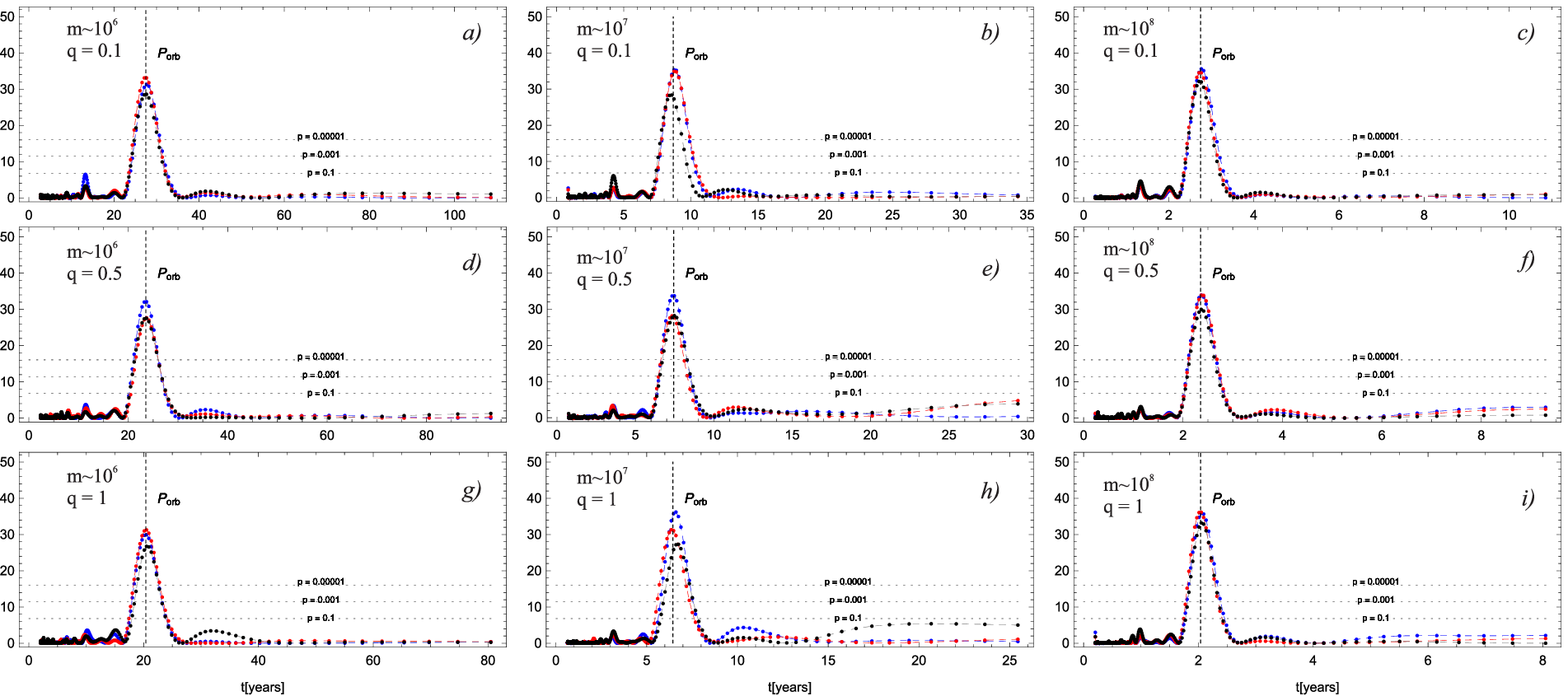}
\caption{Periodograms for the continuum (black dots) and H$\beta$ line segments (red with 30\% and blue with 70\% contribution of the cBLR) for the light curves given in the Fig.
\ref{fig:var_678_01_05_1}.}
\label{fig:lcp_678_01_05_1}
\end{figure*}

Finally, we explore periodicity in a number of simulated systems shown in Fig. \ref{fig:lc_678_01_05_1_per}. For each case from Fig. \ref{fig:lc_678_01_05_1_per}, we calculate the continuum and line emission variability for four complete orbits (shown in Fig.\ref{fig:var_678_01_05_1}). We  assumed two different levels of contribution of the cBLR emission to the total line emission: 30\% (red points in Fig. \ref{fig:var_678_01_05_1}) and 70\% (blue points in Fig. \ref{fig:var_678_01_05_1}). The continuum light curve in Fig. \ref{fig:var_678_01_05_1} is represented with black dots. Their periodograms are plotted in  Fig. \ref{fig:lcp_678_01_05_1}.

Figs. \ref{fig:var_678_01_05_1} and  \ref{fig:lcp_678_01_05_1} show that the periodicity is clearly present in the line and continuum light curves. 
The significance of periodogram peaks is very similar, and the minor difference is due to the applied white and read noise. 
Additionally, we can see that for high mass systems periodogram peaks are slightly higher in absolute value, than in the lower mass case.

\section{Conclusions}
\label{sec:conclusion}

For the first time we present long-term simulations of the expected spectrophotometric variability of SMBBHs, taking different dynamical parameters, and also specific characteristics of the SMBBHs, as e.g. their mass, mass ratio, separation, and eccentricities. In addition, we include the perturbation of the disc temperature  due to the mutual gravitational interaction of the two components and change in the accretion rate of the SMBBH system.
These two effects cause periodical variations in the line and continuum light curves. Here we explore the correlation between the continuum and line variability, and test the possibility of detecting the periodicity from a simulated light curve, measured from the series of modeled spectra  with white and red noise added, to resemble real observations.

\noindent From our investigations we can outline the following conclusions:

\begin{enumerate}[(i)]
\item The model shows that the continuum and line profiles of sub-pc SMBBHs strongly depend on the total mass of the binary system. In the case of very massive SMBBHs ($\sim 10^8\mathrm{M\odot}$), the broad line profile is mostly emitted from the cBLR and does not vary. The broad lines emitted from the BLR1+BLR2 (within Roche lobes, close to the accretion discs) contribute to the continuum around the broad line.
\item We explore variability in the continuum and broad line luminosity and find that the long-term monitoring of SMBBH candidates can give valid information about the nature of these objects, especially in the case of more massive systems for which one can expect larger interaction between SMBBH components (accretion discs - contributing to the continuum emission, and BLRs) and thus larger continuum and line flux variability.
\item The line flux should have periodical variability that indicates the presence of SMBBH. The continuum flux in the case of more massive SMBBHs also shows periodicity. However, the periodicity in the continuum can be hidden by possible limitation in the observational set-up, or brightness of the object. Here we demonstrate the influence of  white and red noise added to the modeled SED, and show that both periodicity in the light curve, as well as some effects in the broad line profiles caused by the dynamics of the binary system, may be affected by the noises in the spectra. Therefore, the high quality spectral observations of the SMBBH candidates should be performed in order to confirm or rule out the binary hypothesis.
\item We found that the H$\beta$ line variability is mostly present in the line core and shows strong correlation with continuum. 
\end{enumerate}

In this paper we give a consistent model of the SMBBH which is based on the relationships between the continuum luminosity, mass of central black hole, and dimensions of the BLR, that introduce some initial constraints to the model. Additionally we explore some effects in the time evolution of the SMBBH and influence of the noise to line profiles, and consequently to the light curves. The effects of different cadence, which is very important for detection of namely the periodicity from light curves, will be studied in the forthcoming paper.

\section*{Acknowledgments}

The authors acknowledge funding provided by the Astronomical Observatory Belgrade (the contract 451-03-68/2020-14/200002), University of Belgrade - Faculty of Mathematics (the contract 451-03-68/2020-14/200002), University of Kragujevac - Faculty of Sciences (the contract 451-03-9/2021-14/200122) through the grants by the Ministry of Education, Science, and Technological Development of the Republic of Serbia. D.I. acknowledges the support of the Alexander von Humboldt Foundation. We thank to the referee for very useful comments.
A. K. and L. C. P. acknowledge the support by Chinese Academy of Sciences President’s International Fellowship Initiative (PIFI) for visiting scientist.

\section*{Data availability}

No new data were generated or analysed in support of this research.

\appendix

\section{Approximation of the disc perturbation}
\label{sec:appendixA}

Here we do not consider semi-detached binary systems, but only the systems which have gravitational interaction,
therefore we assume that the discs are stable but perturbed by tidal perturbation, which can affect the disc temperature.

To find the interaction between two black holes (BHs) and their influence on the disc emission, we use virial theorem, which balanced the gravitational potential ($U$) and kinetic energy ($Q$) as $2Q=U$. We assume that this theorem can be applied to any mass element in the accretion discs of each component.
\begin{figure}
\includegraphics[width=8.5cm]{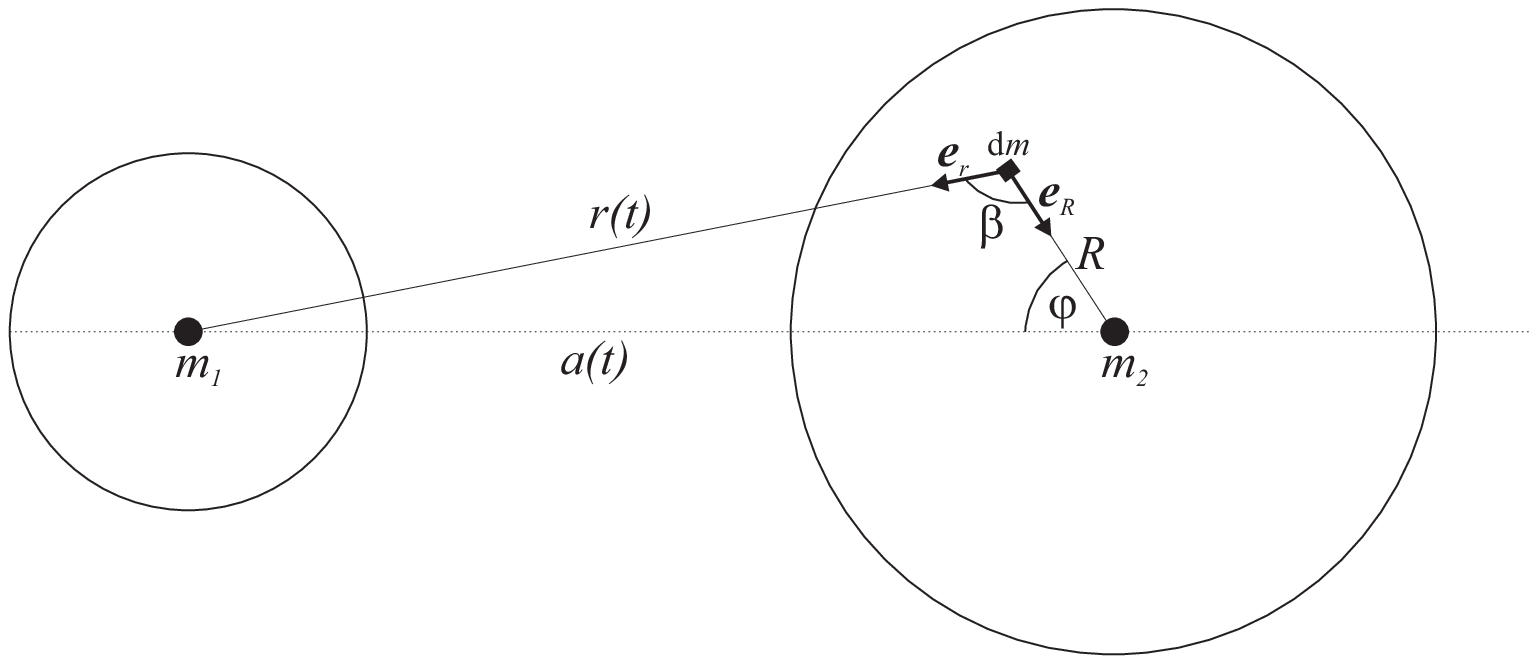}
\caption{The gravitational perturbation experienced by an element of mass ($dm$) in the accretion disc of secondary SMBH due to interaction with primary SMBH.}
\label{fig:sketch_binary}
\end{figure}

Fig. \ref{fig:sketch_binary} shows two BHs in a general configuration with their respectful accretion discs. We select an arbitrary mass element $dm$ in the accretion disc of component 2. The gravitational potential energy of this element within gravitational field of $m_2$ and without perturbation $\delta U$  is:
\begin{equation}
U_{m_2}=G\frac{m_2 dm}{R}
\label{eq-a1}
\end{equation}
where $R$ is distance from the center of the BH with mass of $m_2$, and $G$ is the gravitational constant.

If there is a gravitational force influence of the mass $m_1$ at distance $r$ from the mass element $dm$, the perturbation in the mass element potential ($\delta U$) is caused by the small fraction of pulling the element to the direction of the $m_1$ center. If we define unit vectors $\vec{e}_R$ which follows gravitational force of $m_2$ and unit vectors $\vec{e}_r$ which follows gravitational force of $m_1$  (see Fig. \ref{fig:sketch_binary}) we can assume that the potential energy perturbation is given as follows:
\begin{equation}
\delta U(t)=G\frac{m_1 dm}{r(t)}(\vec{e}_r\cdot\vec{e}_R) = G\frac{m_1 dm}{r(t)}\cos\beta
\label{eq-a2}
\end{equation}
where $\beta$ is the angle between the unit vectors (see Fig. \ref{fig:sketch_binary}).

The total potential energy of the mass element $dm$ is now given as:
\begin{equation}
U_\textrm{tot}=U+\delta U(t)=G\frac{m_2 dm}{R}+ G\frac{m_1 dm}{r(t)}\cos\beta
\label{eq-a3}
\end{equation}

The aim of the paper is not to give the theory of disc perturbation, but to investigate the long-term variability in the line and continuum light curves. However, let us very roughly test validity of the Eq.\ref{eq-a3}, asking for the condition that the $U_\textrm{tot}=0$ in the binary system, which gives:
\begin{equation}
\frac{m_2 dm}{R}=-G\frac{m_1 dm}{r(t)}\cos\beta
\label{eq-a4}
\end{equation}
For $dm$ that is between $m_1$ and $m_2$, i.e., $\cos\beta=-1$ we will have that gravitational potentials of both masses are equal, that is expected in the case that $U_\textrm{tot}=0$.

On the other side, the internal energy of the mass element is directly proportional to the energy that it radiates, i.e.,  to the energy density $\sigma_T T_\mathrm{eff}^4$, and  using the viral theorem we get:
\begin{equation}
2\sigma_T T_\mathrm{eff}^4(t)\Delta t=\left(G\frac{m_2 dm}{R}+G\frac{m_1 dm}{r(t)}\cdot\cos\beta\right)
\label{eq:grav_E_pert1}
\end{equation}
In the case of an unperturbed system, one can write a similar equation:
\begin{equation}
2\sigma_T T_{o}^4\Delta t=G\frac{m_2 dm}{R}
\label{eq:grav_E_unpert}
\end{equation}

Now we find the ratio of Eqs. \ref{eq:grav_E_pert1} and \ref{eq:grav_E_unpert}, that will give temperature profile of a perturbed disc of the component $i=2$, by the component $j=1$) as a function of the unperturbed  disc temperature:
\begin{equation}
T^{î}_\mathrm{eff}(t)=T_{o}\left(1+\frac{m_j}{m_i}\frac{R\cdot\cos\beta}{r(t)} \right)^{1/4}
\label{eq:Teff_pert}
\end{equation}

Eq. \ref{eq:Teff_pert} gives that the effective temperature in the perturbed disc can be smaller ($\cos\beta<0$ for the side oriented to the perturbing component), or larger ($\cos\beta>0$ for another side) from unperturbed temperature across the disc. This can be explained with pulling effect of the perturbing BH, and in that case, the effect of accretion will be smaller than in the case where the pulling effect of perturbing BH is supporting the pulling effect of the central BH.

Considering a disc part with distance $R$ from the center of component $i$, and $r$ from component $j$ (here we use the same notation as in \S 2.3, Eq. \ref{eq:r_od_fi}), it is easy to get from simple geometrical consideration (see Fig. A1) that (see Eq. \ref{eq:grav_E_pert1})
\begin{equation}
r(t)=a(t)\sqrt{1+\left({R\over a(t)}\right)^2-2{R\over a(t)}\cdot\cos(\varphi)},
\label{eq:r_od_t}
\end{equation}
where $a(t)$ is distance between two black holes, and $\varphi$ is the angle between $a$ and $R$ observed from
the center of component $i$.

\section{Change in accretion rate}
\label{sec:appendixB}

In order to include the changes in accretion rates of SMBBH components due to surrounding matter flow to the SMBBH components. This effect should affect the amount of accreating gas around an SMBH in the binary system, and therefore to the accretion rate in this component.  To include this we started that the component $i$ has an accretion rate $f_e$ for single SMBH  given in Eq. (9), and that this can be calculated as
\citep[see Eq. (3) in][]{mi05}:
\begin{equation}
f_{Ei}={3\over 4}{{\nu_i\Sigma_i\mu c\sigma_T}\over {G M_i m_p}},  
\label{eq:accretion}
\end{equation}
where $\mu$ is the viscosity, $\Sigma$ is the disc surface density of component $i$ taken as a constant and computed for outer edge of the accretion discs, $m_p$ is the mass of proton, $\sigma_T$ is the Thomson cross section, $c$ is the speed of light and we take $\mu\sim 0.1$ \citet[][]{mi05}.

To find the accretion rate for a component of binary, similar as in \citet{fa14}, we assumed that each mini disc is an alpha- disc with $h/r$ = 0.1 and $\alpha$ = 0.1, that gives viscosity for component $i$ as
\citep[][see their Eq. 15]{fa14}:
\begin{equation}
\nu_i\approx 2\pi r_i(t)^2\cdot 10^{-3} \left({a(t)\over r_i(t)}\right)^{3/2} \left({m_i\over M}\right)^{1/2}P,  
\label{eq:viscosity}
\end{equation}
where $r_i$ is taken to be distance from the barycentre to the $i$-th component (see Fig. 4) , $a$ is the distance between the components, $M$ is the total mass of binaries and $P$ is the period.

\begin{figure}
    \centering
    \includegraphics[width=8.5cm]{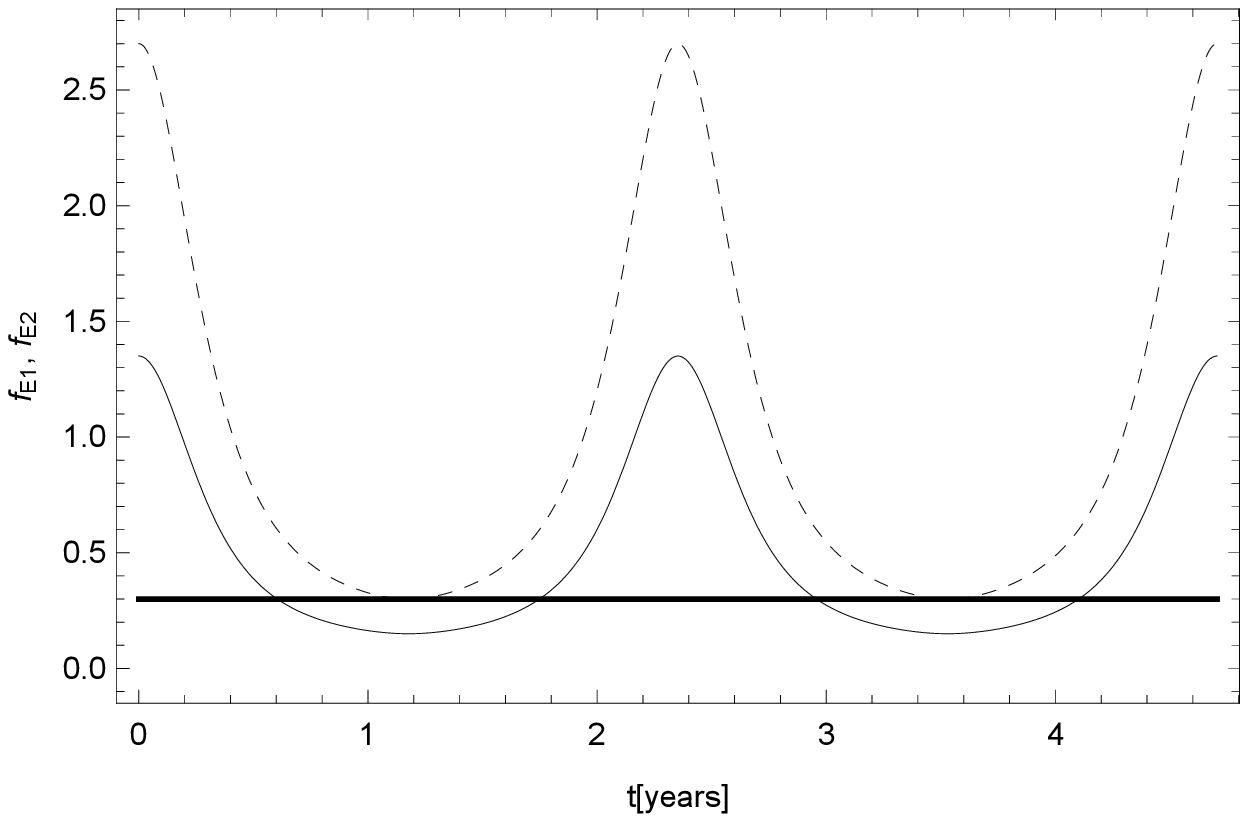}
    \caption{Accretion rate for both component given in Eddington accretion, for middle point on the disc during full orbital rotation. Dot-dashed line represents component 1 and dashed line represent component 2.
    The solid horizontal line represents the accretion rates components in the case single SMBH. Computations has been done for $m_1=5\times10^8M_{\odot}$ and $m_2=10\times10^8M_{\odot}$.}
    \label{fig:fE}
\end{figure}

Using this, we calculated changes in the accretion rate from both components (see Fig. \ref{fig:fE}). Then the final temperature profile has been calculated as:

\begin{equation}
T^i_\mathrm{eff}(t)=T^i_{o}\left(1+\frac{m_j}{m_i}\frac{R\cdot\cos\beta}{r(t)} \right)^{1/4} \cdot \left(\frac{f_{Ei}(t)}{f^{0}_{Ei}}\right)^{1/4},
\label{eq:Teff_pert2}
\end{equation}
where $r(t)$ is connected with distance between components, as it is given in Eq.\ref{eq:r_od_t} and $f^0_{Ei}$ represents scale factor which we adopt as 0.3 for both components. The effective temperature of the component $i$ is time dependent, and therefore the luminosity of this and another component is time dependent. This produce variability in a SMBBH system.

\end{document}